\documentclass[12pt]{iopart}

\usepackage{graphicx,color,subfigure}
\usepackage{bm,bbm}
\usepackage{amssymb}
\usepackage[squaren]{SIunits}
\usepackage{ulem}

\definecolor{mygreen}{rgb}{0,0.5,0} 
\definecolor{myblue}{rgb}{0,0,0.75} 
\definecolor{mymagenta}{cmyk}{0,1,0,0.12}

\newcommand{\cov}{\mathrm{cov}}
\newcommand{\var}[1]{\mathrm{var}(#1)}

\newcommand{\ave}[1]{\ensuremath{\langle#1\rangle}}

\newcommand{\expect}[1]{\left<#1\right>}
\newcommand{\lowexpect}[1]{\left<\right.#1\left.\right>}
\newcommand{\NA}{N_{\rm A}}
\newcommand{\NAb}{\bar{N}_{\rm A}}
\renewcommand{\NL}{N_{\rm L}}
\newcommand{\nl}{n_{\rm L}}
\newcommand{\hj}{\hat{\jmath}}
\newcommand{\hf}{\hat{f}}

\DeclareMathSymbol{\jj}{\mathalpha}{letters}{"11}
\newcommand{\jhat}{\mathbf {\hat{\jj}}}
\newcommand{\fhat}{{\hat{\bf f}}}
\newcommand{\Jhat}{{\hat{\bf J}}}
\newcommand{\Fhat}{{\hat{\bf F}}}
\newcommand{\hS}{\hat{S}}

\newcommand{\Shat}{\hat {\bf{S}}}
\newcommand{\Sbar}{\bar{\bf{S}}}
\newcommand{\hlambda}{\hat{\lambda}}
\newcommand{\hLambda}{\hat{\Lambda}}
\newcommand{\lambdab}{{\boldsymbol \lambda}}
\newcommand{\Lambdab}{{\boldsymbol \Lambda}}
\newcommand{\lambdahat}{{\hat {\boldsymbol \lambda}}}
\newcommand{\Lambdahat}{\hat {\mathbf{\Lambda}}}

\newcommand{\Lambdabar}{\bar{\mathbf{\Lambda}}}

\newcommand{\myV}{V}
\newcommand{\bV}{{\bf{\myV}}}
\newcommand{\Vhat}{\hat{\bf{\myV}}}
\newcommand{\Vbar}{\bar{\bf{\myV}}}
\newcommand{\dVhat}{{\delta \hat{\bf{\myV}}}}

\newcommand{\supi}{^{(i)}}

\newcommand{\supm}{^{(m)}}

\newcommand{\npulse}{N_{\rm pulses}}
\newcommand{\struc}[3]{{c^{#1#2}_{#3}}}
\newcommand{\tGm}{{\bf H}^{(m)}}
\newcommand{\dd}[2]{{\delta_{#1}^{#2}}}
\newcommand{\vnv}{{\rm \hat{V}}}
\newcommand{\vnvb}{{\rm \bar{V}}}

\newcommand{\cm}[1]{{\cal C}(#1)}
\newcommand{\X}{\mathcal{X}}
\newcommand{\p}{p}
\newcommand{\Xsc}{\X}

\newcommand{\Bp}{\stackrel{\leftrightarrow}{\bf B}_1}
\renewcommand{\Bp}{{\mathbb B}_1}
\newcommand{\rhoat}{\rho_{\rm at}}
\newcommand{\bx}{{\bf x}}
\renewcommand{\Bp}{\tilde{\bf B}}
\newcommand{\Aeval}{a}
\newcommand{\Aevec}{{\bf v}}
\newcommand{\decay}{r_i(t)}
\newcommand{\matA}{\mathcal{A}}
\newcommand{\matD}{\mathcal{D}}
\newcommand{\matT}{\mathcal{T}}
\newcommand{\matN}{\mathcal{N}}
\newcommand{\matF}{\mathcal{F}}

\newcommand{\tgauss}{\tau_{\rm gauss}}
\newcommand{\rtthr}{{\sqrt{3}}}
\newcommand{\width}{{w}}
\newcommand{\Bunit}{\hat{B}}
\newcommand{\bBperp}{{\bf B}_{\perp}'}
\newcommand{\bBpar}{{\bf B}_{\parallel}'}
\newcommand{\Bbar}{\bar{\bf{B}}}

\begin{document}
\title{Quantum atom-light interfaces in the gaussian description for spin-1 systems}

\author{Giorgio Colangelo$^{1,*}$, Robert J. Sewell$^{1,*}$, Naeimeh Behbood$^{1}$, Ferran Martin Ciurana$^{1}$ and Morgan W. Mitchell$^{1,2}$}

\address{
$^{1}$ ICFO - Institut de Ciencies Fotoniques, Mediterranean Technology Park, 08860 Castelldefels, Barcelona, Spain\\
\noindent $^{2}$ ICREA - Institucio Catalana de Recerca i Estudis Avancats, 08015 Barcelona, Spain \\
\noindent $^*$ These authors contributed equally to this work.
}

\ead{giorgio.colangelo@icfo.es}

\begin{abstract}
We extend the covariance-matrix description of atom--light quantum interfaces, originally developed for real and effective spin-1/2 atoms, to include ``spin alignment'' degrees of freedom.  This allows accurate modeling of optically-probed spin-1 ensembles in arbitrary magnetic fields.
We also include technical noise terms that are very common in experimental situations.
These include magnetic field noise, variable atom number and the effect of magnetic field inhomogeneities.
We demonstrate the validity of our extended model by comparing numerical simulations to a free--induction decay (FID) measurement of polarized $^{87}$Rb atoms in the $f = 1$ ground state.
We qualitatively and quantitatively reproduce experimental results with all free parameters of the simulations fixed.
The model can be easily extended to larger spin systems, and adapted to more complicated experimental situations.
\end{abstract}

\section{Introduction}
Atomic ensembles play an essential and growing role in quantum optics \cite{HammererRMP2010}, with applications in quantum networking \cite{JulsgaardN2004}, generation of optical quantum resources \cite{MatskoPR2002, MikhailovOL2008}, and quantum-enhanced instruments \cite{GiovannettiPRL2006, LouchetChauvetNJP2010, NapolitanoN2011, SewellPRL2012}. 
The continuous-variable (CV) approach \cite{BraunsteinRMP2005} efficiently describes experiments involving many quanta.  The great majority of CV atomic ensemble experiments are performed with gaussian states, although non-gaussian atomic states \cite{DubostPRL2012} are required for some tasks \cite{GiedkePRA2002}.  Gaussian states can be described very economically in terms of mean values and variances, whereas general states require a description exponential in the size of the system.
Madsen and M{\o}lmer  \cite{MadsenPRA2004} introduced a covariance matrix technique to describe collective spin variables in quantum interfaces between light and spin-1/2 ensembles.  A wide variety of effects, including spatial and temporal inhomogeneities, loss, decoherence, atomic transport and projective measurements have been incorporated into this framework \cite{KoschorreckJPAMOP2009}.  An important omission until now has been the description of larger-spin systems.  Most atomic ensemble experiments are performed with alkali atoms and thus have ground-state spin of at least 1, which implies tensorial light shifts \cite{MadsenPRA2004} and generalized Faraday rotation effects \cite{EchanizPR2008}.  An extension of the covariance matrix techniques to include these will allow statistical description of many large-spin applications, including quantum state characterization \cite{SmithPRL2004} and preparation \cite{Pustelny2011},  quantum chaos \cite{ChaudhuryN2009}, optical magnetometry \cite{SewellPRL2012,BudkerPRL2000, PustelnyJAP2008} and quantum non-demolition measurement \cite{KoschorreckPRL2010b, SewellARX2013}.

While several earlier works have applied the spin-1/2 framework to spin-1 or larger systems through the identification of a two-state ``pseudo spin-1/2''  sub-system, there are scenarios in which the dynamics naturally involves more than two levels, and requires a more expanded description.  A clear example is  a spin-1 or larger atom in the presence of both magnetic and optical fields.  
The magnetic field couples Zeeman states differing by $\Delta m = \pm 1$, whereas the optical fields couple also states with $\Delta m = \pm 2$ through tensorial light shifts.  This system has been much studied using density-matrix approaches \cite{SmithPRL2004}, which describe fully the average single-atom properties but not the noise properties, which arise from correlations among the atoms.  Consideration of the quantum noise in these systems  motivates the current work, in which we extend the covariance matrix approach to spin-1 atoms.  To the suite of techniques available for spin-1/2 ensembles \cite{MadsenPRA2004,KoschorreckJPAMOP2009}, we add the ability to treat both vectorial and tensorial light shifts, technical noise due to uncertainty in the atom number, and dephasing due to magnetic field inhomogeneities.

The paper is  organized as follows: in section \ref{sec:Formalism} we present the formalism, which employs eight orientation and alignment operators to describe the $F=1$ collective atomic spin. In section \ref{sec:Dynamics} we analyze the spin dynamics in the presence of probing light and an external magnetic field, including coherent evolution, decoherence due to scattering of probe photons, and dephasing due to inhomogeneous magnetic fields.  In section \ref{sec:Measurement} we review the description of optical measurement within the covariance matrix formalism.  In section \ref{sec:TechnicalNoise} we describe the initial state including technical noise from uncertain atom number.   In sections \ref{sec:FID} and \ref{sec:Simulation} we compare numerical results of our model with experimental data and identify the practical limits of the gaussian approximation in this system. 

\section{Formalism}
\label{sec:Formalism}

We work with collective operators describing macroscopic numbers of particles, for which a continuous-variable description is appropriate.  
Throughout, we use the covariance matrix techniques~\cite{MadsenPRA2004, KoschorreckJP2009}, which are sufficient to describe the Gaussian states encountered in the great majority of continuous-variable experiments. 

\subsection{Quantum polarization description}
Polarized light in continuous variables can be described with Stokes operators:
\begin{equation}
\label{eq:2}
\hat{S}_x\equiv\frac{1}{2}\hat{\mathbf{a}}^{\dag}\mathbf{\sigma}_x\hat{\mathbf{a}}
\quad \hat{S}_y\equiv\frac{1}{2}\hat{\mathbf{a}}^{\dag}\mathbf{\sigma}_y\hat{\mathbf{a}}
\quad \hat{S}_z\equiv\frac{1}{2}\hat{\mathbf{a}}^{\dag}\mathbf{\sigma}_z\hat{\mathbf{a}}
\end{equation}
Where $\hat{\mathbf{a}}\equiv(\hat a_+,\hat a_-)^T$ and $\hat a_+, \hat a_-$ are the annihilation operators for the left and right circular polarization and $\mathbf{\sigma}_x$, $\mathbf{\sigma}_y$, $\mathbf{\sigma}_z$ the Pauli matrices. 
The $\hat{S}_x$, $\hat{S}_y$ and $\hat{S}_z$ Stokes operators represent respectively linearly polarized light horizontally or vertically, linearly polarized light on the $\pm 45^o$ direction, left and right circularly polarized light. 
They have the same commutation relations as angular momentum operators, $[\hat{S}_x,\hat{S}_y] =i \hat{S}_z$ and cyclic permutations.  

\subsection{Description of spin-1 ensembles}
A single spin-1 atom is described by a density matrix with $8$ degrees of freedom, which we express in terms of 8 single-particle operators $\hat{\lambda}_i$.  
These generalize the Pauli matrices, in the sense that they are traceless, Hermitian, and  obey the orthonormality relation $\rm{Tr}(\lambda_i\lambda_j) = 2\delta_{ij}$.  
The first three operators are the components of the spin vector $\hat{\bf f}$, obeying $[\hf_x,\hf_y] = i \hf_z$.  For illustration, we give the spin-1 matrix representation:
\newcommand{\feq}{\stackrel{F=1}{\longrightarrow}}
\begin{eqnarray}
\nonumber
\hat f_x &\feq& \frac{1}{\sqrt{2}}
\left(
 \begin{tabular}{*{3}{c}}
  0 & 1  & 0 \cr
  1 & 0 &  1 \cr
  0 & 1  & 0
 \end{tabular} \right),
\\ \nonumber
\hat f_y &\feq& \frac{1}{\sqrt{2}}
\left(
\begin{tabular}{*{3}{c}}
  0 & -i  & 0 \cr
  i & 0 &  -i \cr
  0 & i  & 0
 \end{tabular} \right),
\\ \nonumber
\hat f_z &\feq&
\left(
 \begin{tabular}{*{3}{c}}
  1 & 0  & 0 \cr
  0 & 0 &  0 \cr
  0 & 0  & -1
 \end{tabular} \right)\,.
\end{eqnarray}
The others are rank-2 tensor operators, for which we use the symbol $\jhat$, with components (again with the spin-1 representation for illustration):  
\begin{eqnarray}
\nonumber
\hj_x \equiv \hat f_x^2-\hat f_y^2 \feq \left(
 \begin{tabular}{*{3}{c}}
  0 & 0  & 1 \\
  0 & 0 &  0 \\
  1 & 0  & 0
 \end{tabular} \right), 
\\ \nonumber
 \hj_y \equiv\hat f_x\hat f_y+\hat f_y\hat f_x \feq \left(
 \begin{tabular}{*{3}{c}}
  0 & 0  & -i \cr
  0 & 0 &  0 \cr
  i & 0  & 0
 \end{tabular} \right), 
\\ \nonumber
 \hj_k \equiv\hat f_x\hat f_z+\hat f_z\hat f_x \feq \frac{1}{\sqrt{2}}
\left(
 \begin{tabular}{*{3}{c}}
  0 & 1  & 0 \cr
  1 & 0 &  -1 \cr
  0 & -1  & 0
  \end{tabular} \right), 
\\ \nonumber
\hj_l \equiv\hat f_y\hat f_z+\hat f_z \hat f_y \feq \frac{1}{\sqrt{2}}
\left(
 \begin{tabular}{*{3}{c}}
  0 & -i  & 0 \cr
  i & 0 &  i \cr
  0 & -i  & 0
 \end{tabular} \right),  
\\ \nonumber
 \hj_m\equiv\frac{1}{\sqrt{3}}(2\hat f_z^2-\hat f_x^2-\hat f_y^2) \feq \frac{1}{\sqrt{3}}
\left(
 \begin{tabular}{*{3}{c}}
  1 & 0  & 0 \cr
  0 & -2 &  0 \cr
  0 & 0  & 1
 \end{tabular} \right)
\label{eq:jdefs}
\end{eqnarray}
With quantization axis along $z$, $\hj_m$ describes the population imbalance between $m_F = 0$ and other states, while $\hj_{x,y}$  describe $m_F = \pm 1$ coherences.  $\hj_{k,l}$ represent  $m_F = \pm 1$ coherences in other quantization axes.

We note that the above operator definitions are spin-independent, and that the results in this manuscript follow from these operator definitions, not from the spin-1-specific matrix representations.  
As we shall see below, the most important coherent interactions: Larmor precession, Faraday rotation, and tensorial light shifts, can be fully described using the above operators, even for larger spin.  
The formalism developed here is thus applicable to some scenarios involving spin-3/2 and higher.  
Not all processes can be explained using just $\fhat$ and $\jhat$ operators, however.  
For example, with spin-2 atoms modulated optical pumping in the presence of a B-field has been used to produce hexadecapole moment due to coherence between Zeeman states with $\Delta m_F = 4$~\cite{AcostaOE2008}. 

\begin{figure}
 \centering
\includegraphics[width=\textwidth]{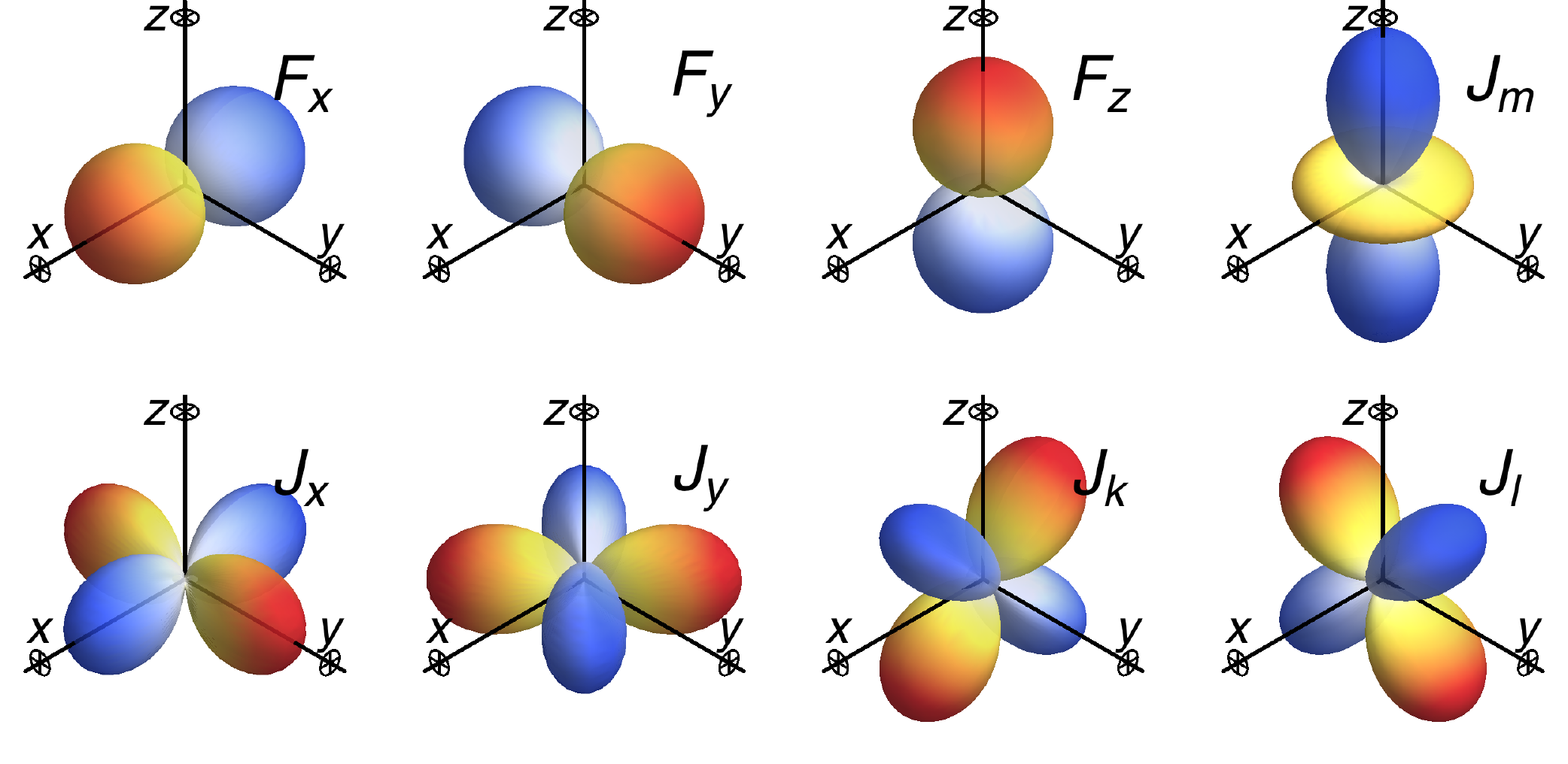}
\caption{Contribution $dW/d\bar{\lambda}_i$ of the $f$ and $j$ operators to the Wigner distribution $W_\rho(\theta,\phi)$ representing the state $\rho =\frac{1}{3} {\bf I} + \frac{1}{2} \sum_i \hlambda_i \bar{\lambda}_i$.  Radius indicates magnitude, warm (cold) colors indicate positive (negative) contributions.  Axis markers indicate unity.  }
\label{fig:DistributionVisualization}
\end{figure}

\subsection{Spin visualization}
From the orthogonality relation $\rm{Tr}[\lambda_i\lambda_j] = 2\delta_{ij}$, an arbitrary single-atom density matrix $\rho$ can be expressed as 
\begin{equation}
 \rho = \frac{1}{3} {\bf I} + \frac{1}{2} \sum_i \hlambda_i \bar{\lambda}_j 
\end{equation}
where $ \bar{\lambda}_i \equiv {\rm Tr}[\rho \hlambda_i]$.  This suggests a visualization in terms of the spin Wigner distribution $W(\rho)$, which is efficiently calculated as in Dowling, et al.  \cite{DowlingPRA1994}.  In Figure \ref{fig:DistributionVisualization} we show the differential contribution to the Wigner distribution $d W / d  \bar{\lambda}_i$.  This shows, for example, that $\hj_x,\hj_y,\hj_k$ and $\hj_l$ are related by spatial rotations while $\hj_m$ is distinct.

\subsection{Commutation relationships}
The operators $\fhat$, $\jhat$, and $\Shat$ have commutators given by
\begin{equation}
  [\hlambda_a,\hlambda_b] = i \struc{\hlambda_a}{\hlambda_b}{\hlambda_k} \hlambda_k
\label{commCompact}
\end{equation}
where the $\hlambda$ are $\fhat$ or $\jhat$ components and a sum is over $k$ is implied.  
The structure constants $\struc{\hlambda_a}{\hlambda_b}{\hlambda_k}$ are completely antisymmetric in the three indexes, and 
\begin{eqnarray}
\struc{\hf_x}{\hf_y}{\hf_z}&=& 1 , \quad
\struc{\hj_x}{\hj_y}{\hf_z}  =   2 , \quad  \struc{\hf_x}{\hj_l}{\hj_m} = \struc{\hf_y}{\hj_m}{\hj_k} = \sqrt{3},  \\
        \struc{\hf_x}{\hj_y}{\hj_k} &=& \struc{\hf_x}{\hj_l}{\hj_x}=\struc{\hf_y}{\hj_k}{\hj_x} = \struc{\hf_z}{\hj_k}{\hj_l} = 1.  \label{eq:StructureFactorsLambda}
\end{eqnarray}
To this we can add
\begin{eqnarray}
\struc{\hS_x}{\hS_y}{\hS_z}&=& 1  \label{eq:StructureFactorsS}
\end{eqnarray}
All structure constants not given above are zero.  
The commutators are given explicitly in Table \ref{tab:comms}.

\begin{table}
\centering
\caption{\label{tab:comms}Commutation relationships for single-atom operators.}
\footnotesize\rm
\begin{tabular}{c|cccccccc}
\br
$[\ ,\ ]$  & $\hf_x$  & $\hf_y$  & $\hf_z$  & $\hj_x$  & $\hj_y$   & $\hj_k$  & $\hj_l$ & $\hj_m$ \\
\mr
$\hf_x$ & 0 & $\hf_z$ & $-\hf_y$ & $-\hj_l$ & $\hj_k$ & $-\hj_y$ & $\rtthr\hj_m+\hj_x$ & $-\rtthr\hj_l$ \\
$\hf_y$ &  $-\hf_z$ & 0 &$\hf_x$ &  $-\hj_k$ & $-\hj_l$ & $-\rtthr\hj_m+\hj_x$ & $\hj_y$ & $\rtthr\hj_k$ \\
$\hf_z$ & $\hf_y$ & $-\hf_x$ & 0 & $2\hj_y$ & $-2 \hj_x$ & $\hj_l$ & $-\hj_k$ &$0$\\
$\hj_x$ & $\hj_l$ & $\hj_k$ & $-2\hj_y$ &  0 &$2\hf_z$ & $-\hf_y$ & $\hf_x$ & $0$ \\
$\hj_y$ & $-\hj_k$ & $\hj_l$ & $2 \hj_x$ & $-2\hf_z$ & 0 & $\hf_x$  & $-\hf_y$ & $0$ \\
$\hj_k$ & $\hj_y$ & $\rtthr\hj_m-\hj_x$ & $-\hj_l$ & $\hf_y$ & $-\hf_x$ & 0 & $\hf_z$  & $-\rtthr\hf_y$ \\
$\hj_l$ &  $-\rtthr\hj_m-\hj_x$ & $-\hj_y$ & $\hj_k$ & $\hf_x$ & $\hf_y$ & $-\hf_z$ & 0 & $\rtthr\hf_x$ \\
$\hj_m$ &    $\rtthr\hj_l$ & $-\rtthr\hj_k$ & 0 & 0 & 0 &$\rtthr\hf_y$ & $-\rtthr\hf_x$ & 0 \\  
\br
\end{tabular}
\end{table}

\subsection{Collective spin operators}
To describe the ensemble  we  define collective operators.  
If  $\lambdahat\supi$ describes atomic operators acting the $i$'th of $\NA$ atoms, then $\Lambdahat \equiv \sum_i^{\NA} \lambdahat\supi$.  
Explicitly for the vector ($\hat {\bf f}$) and tensor ($\jhat$) collective spin operators:
\begin{eqnarray}
\label{eq:5}
\hat{\bf F}\equiv\sum_{i=1}^{\NA}\fhat^{(i)}, \quad 
\hat{\bf J}\equiv\sum_{i=1}^{\NA}\jhat^{(i)}.
\end{eqnarray}
We note that these inherit their commutation relations from the microscopic operators:
$[\hLambda_a,\hLambda_b] = i \struc{\hlambda_a}{\hlambda_b}{\hlambda_k} \hLambda_k$.  
Finally, we define a phase-space vector to describe the state of the whole system
\begin{eqnarray}
\Vhat = {\bf B} \oplus {\Fhat} \oplus {\Jhat} \oplus \bigoplus_{i=1}^{\npulse} {\Shat}\supi 
\label{eq:stateVector}
\end{eqnarray}
where $\oplus$ indicates the direct sum and ${\bf B}$ is the magnetic field vector at the location of the ensemble.  
It should be noted that ${\bf B}$ is here a classical field, whereas the other components of $\Vhat$ are operators.  
The inclusion of ${\bf B}$ in  ${\Vhat}$ allows for classical uncertainty about the field to be incorporated in a natural way into the calculations~\cite{MadsenPR2004}, as we describe below.

\subsection{Covariance matrix}
We work within the gaussian approximation, i.e., we assume that $\Vhat$ is fully characterized by its average $\lowexpect{\Vhat}$ and by its covariance matrix  $\Gamma_\bV$:
\begin{eqnarray}
\label{GammaV}
 \Gamma_\bV\equiv\frac{1}{2}\langle\Vhat\wedge\Vhat+(\Vhat\wedge\Vhat)^T\rangle-\langle\Vhat\rangle\wedge\langle\Vhat\rangle
\end{eqnarray}
where $\wedge$ indicates the outer product.  
It will be convenient to define $\dVhat \equiv \Vhat - \lowexpect{\Vhat}$, the fluctuations of $\Vhat$ about the mean, from which $\Gamma_\bV=\frac{1}{2}\langle\dVhat\wedge\dVhat+(\dVhat\wedge\dVhat)^T\rangle-\langle\dVhat\rangle\wedge\langle\dVhat\rangle$.

\section{Dynamics}
\label{sec:Dynamics}

We describe the most important dynamical effects for light-matter interfaces, namely the light-matter interaction that occurs when a pulse of probe light passes the ensemble, and the rotation due to a magnetic field.  Both of these interactions produce coherent rotations, loss of coherence, and addition of noise.  

\subsection{Light-atom interactions}

The light-atom interaction is described by an effective Hamiltonian which describes the dispersive effects of the electric dipole interaction in second order~\cite{GeremiaPRA2006,EchanizJOQSO2005,Deutsch2010}.
In simulations and in practice, it is very convenient to employ a train of optical pulses for probing the ensemble. 
Even for continuous probing it is useful to treat the probe as a train of contiguous pulses, as this allows a course-grained description of the polarization evolution.
During the time the $m$'th pulse is passing through the ensemble, the effective Hamiltonian is
\begin{eqnarray}
 H_{\mathrm{eff}}\supm=G_{1}\frac{\hat S_{z}\supm}{\tau }\hat F_{z}+G_{2}(\frac{\hat S_{x}\supm}{\tau }\hat J_{x}+\frac{\hat S_{y}\supm}{\tau }\hat J_{y}) 
\label{Heff}
\end{eqnarray}
where  $G_1$ and $G_2$ are coupling constants that depend on the geometry of the atomic ensemble and probe beam, the atomic structure, and the detuning from resonance~\cite{EchanizJO2005}.
We  label the Stokes operators as $\Shat\supm(t)$, $m=1\ldots \npulse$ where $t_m$ indicates the time of arrival at the ensemble.  
Thus $\Shat\supm(t < t_m )$ describes the polarization of the $m$'th pulse before entering the ensemble, while  $\Shat\supm(t > t_m + \tau  )$, where $\tau $ is the pulse duration, describes the polarization of the same pulse after exiting the ensemble.
At time $t_m+\tau $, the $m$'th pulse has left the ensemble (we assume the transit time is much less than $\tau $) and the change in the system is described by 
\begin{eqnarray}
\Lambdahat(t_m+\tau ) &=&\Lambdahat(t_m)-i \tau  [\Lambdahat(t_m),H_{\mathrm{eff}}\supm] 
\label{eq:EvOp}
\end{eqnarray}
and
\begin{eqnarray}
{\Shat}\supm(t_m + \tau ) & = & {\Shat}\supm(t_m)  -i\tau [ {\Shat}\supm(t_m),H_{\mathrm{eff}}\supm].
\label{eq:EvS}
\end{eqnarray}
All other polarizations $\Shat^{(n)}, n \ne m$ are unchanged.  
These first-order difference equations will be accurate for sufficiently low-energy pulses, i.e., for small $G_{1,2} \left<\right.{\Shat}\left.\right>$.
For any given physical situation these conditions can be satisfied by subdividing long or high-energy pulses into sub-pulses with smaller $\tau $, at the cost of additional computation time.  
Similarly, if the full ensemble produces large rotations of $\Shat$, the $\Lambdahat$ can be subdivided as in~\cite{KoschorreckJP2009}.  
In the simulations described below we subdivide $\Shat$ until the results converge.

The evolution is compactly expressed in terms of a tensor $\tGm$ containing the coupling constants $G_{1,2}$ and $g_F \mu_0$ and the structure factors $\struc{\vnv_i}{\vnv_j}{\vnv_k}$:
\begin{eqnarray}
\label{eq:vEvolutionTensor}
\vnv_i (t_m +\tau ) &=& \vnv_i(t_m)  +  \vnv_j(t_m)  H_{jk}^{i(m)}  \vnv_k(t_m) \quad\mbox{where}  \\
\label{eq:GTensorDef}
H_{jk}^{i(m)} &\equiv& G_1 \dd{\vnv_j}{\hS_z^{(m)}} \struc{\vnv_i}{\hf_z}{\vnv_k}+G_2 \left( \dd{\vnv_j}{\hS_x^{(m)}} \struc{\vnv_i}{\hj_x}{\vnv_k}+\dd{\vnv_j}{\hS_y^{(m)}} \struc{\vnv_i}{\hj_y}{\vnv_k}\right)
\end{eqnarray}
where $\dd{\vnv_i}{\vnv_j}$ is 1 for $\vnv_i=\vnv_j$ and 0 for $\vnv_i\ne\vnv_j$, and we assume summation over repeated indices.

\subsection{Linearization}

The difference equations Eq.~(\ref{eq:vEvolutionTensor}) are bilinear in the components of $\Vhat$.  
Although nonlinearity can in some cases lead to non-gaussian phase-space distributions~\cite{ChaudhuryN2009,JuliaDiazPRA2012}, in practice gaussian or near-gaussian distributions are far more common, and indeed producing measurably non-gaussian distributions is non-trivial~\cite{DubostPRL2012}.  
This motivates a linearization of the above equations.
Symbolically we write
\begin{eqnarray}
\label{eq:vhatEvolution}
\Vhat (t_m+\tau ) &=&\Vhat(t_m)  +  \Vhat(t_m) \cdot \tGm \cdot \Vhat(t_m)  
\end{eqnarray}
and separate $\Vhat$ into the average $\Vbar$ and the fluctuations $\dVhat$:
\begin{eqnarray}
\Vhat(t_m+\tau ) & = & 
\Vbar (t_m+\tau ) + \dVhat(t_m+\tau ) \nonumber \\
&=& \Vbar (t_m) +  \Vbar(t_m) \cdot \tGm \cdot \Vbar(t_m)  + \dVhat (t_m) \nonumber \\ 
& & +  \Vbar(t_m) \cdot \tGm \cdot \dVhat(t_m)   +  \dVhat(t_m) \cdot \tGm \cdot \Vbar(t_m)  \nonumber \\ 
& & +  \dVhat (t_m)\cdot \tGm \cdot \dVhat(t_m) 
\end{eqnarray}
When the last term can be neglected, the dynamics reduce to
\begin{eqnarray}
\label{eq:vbarEvolution}
\Vbar (t_m+\tau )  &=& \Vbar (t_m) +  \Vbar(t_m) \cdot \tGm \cdot \Vbar(t_m) \\
		&=& \Vbar (t_m) +  {\bf U}^{(m)}  \cdot \Vbar(t_m)
\end{eqnarray}
where ${\bf U}^{(m)}\equiv\Vbar(t_m) \cdot \tGm$, which describes a nonlinear evolution of the average $\Vbar$, and
\begin{eqnarray}
\dVhat(t_m+\tau ) &=& \dVhat (t_m)
 +   \Vbar(t_m) \cdot \tGm \cdot \dVhat(t_m)  \nonumber \\ 
 & & +   \dVhat(t_m) \cdot \tGm \cdot \Vbar(t_m) \nonumber \\ 
 &  = &  {\bf T}^{(m)} \cdot \dVhat(t_m)   
 \end{eqnarray}
 which describes a linear evolution of the fluctuations $\dVhat$ in terms of the matrix
 \begin{eqnarray}
 \label{eq:TTensorDef}
T_{ik}^{(m)} & \equiv & \dd{i}{k} + \vnvb_j H_{jk}^{i(m) + H_{kj}^{i(m)}} \vnvb_j   
\end{eqnarray}
The covariance matrix $\Gamma$ evolves as
 \begin{eqnarray}
 \label{eq:GammaEvolution}
{\Gamma}(t_m + \tau )  & = & {\bf T}^{(m)} \cdot \Gamma(t_m) \cdot {[{\bf T}^{(m)}]^T}. 
 \end{eqnarray}

\subsection{Optically-induced decoherence }

The above coherent rotations are necessarily accompanied by spontaneous scattering of photons, which produces decoherence of the atomic state \cite{KoschorreckJP2009,EchanizJOQSO2005}.   
An accurate treatment is possible \cite{VasilyevJPB2012}, but requires a detailed accounting of the scattering channels.  
Here we describe a simple noise model, based on that of  Madsen and M{\o}lmer~\cite{MadsenPRA2004}, which  agrees reasonably well with more sophisticated models~\cite{VasilyevPC2012}.

We treat the decoherence as removal of a fraction $1-\X$ of the $\NA$ atoms, followed by the re-addition of a fraction $\p$ of the removed atoms, with random polarizations.  
A common source of ``loss'' of atoms from the system is optical pumping into dark states \cite{EchanizJOQSO2005,VasilyevPC2012}.  
In this article we take $\p=1$, i.e., no loss, which gives an upper bound on the introduced decoherence.
As demonstrated in~\ref{app:decoherence}, this alters the phase space distribution as
\begin{eqnarray}
\label{eq:vbardecohere}
\Lambdabar &\rightarrow &  \X \Lambdabar  \\
\label{eq:GammaFromDecoherenceAgain}
\Gamma_{\bf \Lambda} &\rightarrow &  \X^2\Gamma_{\bf \Lambda}  + \X(1-\X)  \NA  \Gamma_\lambdab + \p (1-\X) \frac{2}{3} \NA \mathbbm{1}_{8\times8}
\end{eqnarray}
where $\Lambdabar \equiv \langle \hat{\Lambda} \rangle $, $\Gamma_{\bf \Lambda} \equiv \frac{1}{2}\langle\Lambdahat\wedge\Lambdahat+(\Lambdahat\wedge\Lambdahat)^T\rangle-\langle\Lambdahat\rangle\wedge\langle\Lambdahat\rangle$ is the atomic part of the covariance matrix $ \Gamma_\bV$, $\Gamma_\lambdab \equiv \frac{1}{2}\langle\lambdahat\wedge\lambdahat+(\lambdahat\wedge\lambdahat)^T\rangle-\langle\lambdahat\rangle\wedge\langle\lambdahat\rangle$ is the single-atom covariance matrix, and the expectations are taken with respect to the average single-atom state $\rho$.  
$\Gamma_\lambdab$ can be found from $\Vbar$ as
\begin{eqnarray}
\Gamma_\lambdab &=& \sum_k \bar{\lambda}_k {\bf M}^{(k)}, \nonumber \\
M^{(k)}_{ij} &\equiv&  \frac{1}{4} \left(  {\rm Tr}[ \hat{\lambda}_k \{ \hat{\lambda}_i ,  \hat{\lambda}_j \} ] -  {\rm Tr}[\hat{\lambda}_k  \hat{\lambda}_i]{\rm Tr}[\hat{\lambda}_k  \hat{\lambda}_j] \right)
\end{eqnarray}
where $\bar{\boldsymbol \lambda} \equiv \lowexpect{\boldsymbol \lambda}_\rho$ is the single-atom average and $\{\cdot,\cdot\}$ indicates the anti-commutator.

The transformation of the atomic covariance matrix due to optically induced decoherence is then
\begin{eqnarray}
 \label{eq:GammaAtomsEvolutionOpticalDecoherence}
{\Gamma}_{\bf \Lambda}(t_m+\tau)  & = & {\bf \mathcal{D}}_{\bf \Lambda}^{(m)} \cdot \Gamma_{\bf \Lambda}(t_m+\tau) \cdot [\bf \mathcal{D}_{\bf \Lambda}^{(m)}]^T + {\bf \mathcal{N}_{\bf \Lambda}^{(m)}}
\end{eqnarray}
where $\Gamma_{\bf \Lambda}(t_m+\tau)$ is given by Eq.~(\ref{eq:GammaEvolution}) and
\begin{eqnarray}
\label{eq:DecoherenceOperatorOpticalDecoherence}
\label{DL}{\bf \mathcal{D}_{\bf \Lambda}^{(m)}} &\equiv \Xsc \mathbbm{1}_{8\times8} \quad\mbox{and} \\
\label{eq:NoiseOperatorOpticalDecoherence}
{\bf \mathcal{N}_{\bf \Lambda}^{(m)}} &\equiv  \X(1-\X)  \NA  \Gamma_\lambdab + \p (1-\X) \frac{2}{3} \NA \mathbbm{1}_{8\times8}.
\end{eqnarray}	

The fraction of atoms $1-\Xsc$ that experience incoherent scattering of probe photons during a single pulse of duration $\tau $ is given by $\Xsc=\exp(-\eta_{\gamma}\nl)$, where $\eta_{\gamma}$ is the probability for an atom to scatter a photon and $\nl$ is the number of photons in the pulse.
A similar decoherence applies to the optical polarization $\hS\supm$ \cite{KoschorreckJP2009}, where the fraction of photons scattered is $1-\epsilon$ and $\epsilon=\exp(-\eta_{\gamma}\NA)$, leading to decoherence and noise terms
\begin{eqnarray}
\label{DS}{\bf \mathcal{D}_{\bf S}^{(m)}} &\equiv  \epsilon\mathbbm{1}_{3\times3} \quad\mbox{and} \\
\label{NS}{\bf \mathcal{N}_{\bf S}^{(m)}} &\equiv  \epsilon(1-\epsilon)  \nl  \Gamma_{\rm S},
\end{eqnarray}
where $\Gamma_{\rm S}=(1/4)\mathbbm{1}_{3\times3}$ is the single photon covariance matrix.
Because scattered photons are lost from the pulse, there is no photonic analogue of the last term of Eq.~(\ref{eq:NoiseOperatorOpticalDecoherence}).   
In the scenarios described below the decoherence of the optical pulses is negligible, because the number of photons is much larger than the number of atoms, and we set $\epsilon=1$.

The full covariance matrix $\Gamma$ then evolves as
 \begin{eqnarray}
 \label{eq:GammaEvolutionOpticalDecoherence}
{\Gamma}(t_m + \tau )  & = & {\bf D}^{(m)} \cdot {\bf T}^{(m)} \cdot \Gamma(t_m) \cdot {[{\bf T}^{(m)}]^T} \cdot {[{\bf D}^{(m)}]^T}  + {\bf N}^{(m)}
 \end{eqnarray}
where ${\bf D}^{(m)}\equiv \mathbbm{1}_{3\times3}\oplus\mathcal{D}_{\bf \Lambda}^{(m)}\oplus{\bf \mathcal{D}_{\bf S}^{(m)}}$ and ${\bf N}^{(m)}\equiv \mathbbm{O}_{3\times3}\oplus\mathcal{N}_{\bf \Lambda}^{(m)}\oplus{\bf \mathcal{N}_{\bf S}^{(m)}}$.

\subsection{Atom-field interaction}\label{sub:Atom-field interaction}
In contrast to the atom-light interaction, the interaction of the atoms with the magnetic field is purely linear, so that  precession by large angles can be described in a single step.  
At the same time, inhomogeneities in the magnetic field introduce dephasing, which requires a different description than given above for scattering-induced decoherence.\footnote{Here we assume that the magnetic fields are static, but time-varying fields can also be modeled, as described in Ref.~\cite{Petersen2006}.}

To describe these effects, we first split the field into homogeneous and inhomogeneous parts as ${\bf B}({\bf x}) = {\bf B}_0 +\Bp(\bx)$, with the assumption that  $|{\bf B}_0| \gg |\Bp(\bx)|$ for $\bx$ within the atomic cloud.
Specifically, if $\rhoat(\bx)$ is the atomic density, we take ${\bf B}_0 = \int d^3 x \, \rhoat(\bx) {\bf B}({\bf x})$, the average over the atoms.
 
Local rotation under ${\bf B}({\bf x})$ can be described by the unitary operator
\begin{equation}
U(\bx, t) = \exp[i \gamma t  |B(x)| \hat{f}_{\bf B}(x)]
\end{equation}
where $\hat{f}_{\bf B}(x)$ is the component of ${\Fhat}$ parallel to the local field and $\gamma = \mu_B g_F / \hbar$ is the gyromagnetic ratio, where $g_F$ is the Land\'e factor and $\mu_B$ is the Bohr magneton.
We note that $\hat{f}_{\bf B}^{n+2}(x) = \hat{f}_{\bf B}^n(x)$ for $n\ge 1$, (true for any $F=1$ spin component), so that a Taylor expansion of $U$ gives
\begin{equation}
U(\bx, t) = \mathbbm{1} + \hat{f}_{\bf B}(x) \sin(\gamma  |B(x)|  t) + \hat{f}_{\bf B}^2(x)[ \cos(\gamma  |B(x)|  t) -1]
\end{equation}
We can now see a qualitative difference between ${\bf B}_\parallel(\bx)$, the component of $\Bp(\bx)$ parallel to ${\bf B}_0$, and ${\bf B}_\perp(\bx)$ the perpendicular components. The effect of ${\bf B}_\parallel(\bx)$ is to change $|{\bf B}|$ and thus the precession frequency, causing a deviation from the behaviour under ${\bf B}_0$ that accumulates with time.  
In contrast, ${\bf B}_\perp(\bx)$ principally changes the direction of $\hat{f}_{\bf B}$, a non-accumulating effect.  
For this reason we focus on the effects of ${\bf B}_\parallel(\bx)$.

Using the Heisenberg equations of motion and the commutation relations  of Eq. (\ref{commCompact}), we find dynamical equations 
\begin{equation}
\frac{d}{dt} \lambdahat(\bx) = - \mu_B g_F |{\bf B}(\bx)| \matA(\bx) \cdot \lambdahat(\bx)
\label{eq:dynamicalFieldEquations}
\end{equation}
where 
\begin{equation}
\label{eq:matA}
\fl
\qquad \matA \equiv \left(
\begin{array}{cccccccc}
\cdot &-b_z &b_y &\cdot &\cdot &\cdot &\cdot &\cdot \\
b_z &\cdot &-b_x &\cdot &\cdot &\cdot &\cdot &\cdot \\
 -b_y&b_x &\cdot &\cdot &\cdot &\cdot &\cdot &\cdot \\
\cdot &\cdot &\cdot &\cdot &-2b_z &b_y &b_x &\cdot \\
\cdot &\cdot &\cdot &2 b_z&\cdot &-b_x &b_y &\cdot \\
\cdot &\cdot &\cdot &-b_y&b_x &\cdot &-b_z & \sqrt{3} b_y\\
\cdot &\cdot &\cdot & -b_x&-b_y &b_z &\cdot &-\sqrt{3} b_x \\
\cdot &\cdot &\cdot &\cdot &\cdot &-\sqrt{3} b_y &\sqrt{3} b_x &\cdot 
\end{array}
\right)
\end{equation}
is a block-diagonal matrix and ${\bf b} \equiv {\bf B}/|{\bf B}|$.  
The equation is solved by
\begin{equation}
\lambdahat(\bx,t) = \sum_i e^{- \omega(\bx) \Aeval_i t} P_i \cdot \lambdahat(\bx,0)
\label{eq:dynamicalFieldSolutions}
\end{equation}
where $\omega(\bx) = \mu_B g_F |{\bf B}(\bx)|$, $\Aeval_i, \Aevec_i$ are the eigenvalues and eigenvectors of $\matA$ with corresponding projectors $P_i \equiv \Aevec_i \wedge \Aevec_i$.  
The eigenvectors are imaginary integers $\{\Aeval\} = i \{-2,-1,-1,0,0,1,1,2\}$, indicating the half-period behaviour of some $\jhat$ variables.
The macroscopic operators evolve as 
\begin{eqnarray}
\Lambdahat(t) &=& \int d^3 x \, \rhoat(\bx)   \sum_i e^{- \omega(\bx) \Aeval_i t} P_i \cdot \lambdahat(\bx,0) \\ 
& = &   \sum_i  e^{- \omega_0 \Aeval_i t}  \int d^3 x \, \rhoat(\bx)  e^{- \tilde{\omega}(\bx) \Aeval_i t} P_i \cdot  \lambdahat(\bx,0)
\end{eqnarray}
where $\omega(\bx) = \omega_0 + \tilde{\omega}(\bx)$.  
We now assume that $\lambdahat(\bx,0)$ is uniform~\footnote{This assumption is clearly violated in some cases, for example in spin-echo experiments.  However, it works remarkably well in many other situations.}, 
so that the $\bx$-dependent terms can be collected as $\decay \equiv  \int d^3 x \, \rhoat(\bx) e^{- \tilde{\omega}(\bx) \Aeval_i t}$.  
This gives 
\begin{eqnarray}
\label{eq:DynamicsUnderB}
\Lambdahat(t) & = &   \sum_i  e^{- \omega_0 \Aeval_i t} \decay P_i \cdot  \Lambdahat(0)
\nonumber \\
& = &  {\bf \matD}_{\bf B}(t) \cdot {\bf \matT}_{\bf B}(t)  \cdot  \Lambdahat(0)
\end{eqnarray}
where $ {\bf \matT}_{\bf B}(t) = \sum_i  e^{- \omega_0 \Aeval_i t}  P_i $ describes the coherent evolution and $ {\bf \matD}_{\bf B}(t) = \sum_i \decay  P_i $ describes the dephasing.

In many situations it is reasonable to assume a Lorentzian distribution for $\tilde{\omega}$.  
For example, in the highly-elongated trap described below, we observe an atomic density $\rho(z)$ well approximated by a Lorenzian $\rho(z) =  \width / \pi(z^2 + \width^2)$ where $\width \approx 48$ mm is the full-width half-maximum (FWHM) extent of the ensemble. 
The decay term is then $\decay =e^{- \width \gamma |\Aeval_i \partial B_\parallel/\partial z|   t }$ or $\decay =e^{- t/T_i }$ with $1/T_i = \width \gamma |\Aeval_i \partial B_\parallel/\partial z|$, and
\begin{eqnarray}
\label{eq:ExponentialDephasing}
{\bf \matD}_{\bf B}(t)  &=&  \sum_{i} e^{-t/T_i} P_i 
\end{eqnarray}

To preserve the uncertainty principle in the presence of the decay implied by Eq. (\ref{eq:ExponentialDephasing}), we must introduce a noise contribution ${\bf N}(t)$ to the covariance matrix \cite{KoschorreckJPAMOP2009}.  
The transformation due to evolution under an inhomogeneous field is thus
\begin{eqnarray}
 \label{eq:GammaEvolutionB}
{\Gamma}_{\bf \Lambda}(t)  & = & {\bf \matD}_{\bf B}(t) \cdot{\bf \matT}_{\bf B}(t) \cdot \Gamma_{\bf \Lambda}(0) \cdot {{\bf \matT}_{\bf B}^T(t)} \cdot {{\bf \matD}^T_{\bf B}(t)} + {\bf \matN}(t)
\end{eqnarray}
where the added noise obeys
\begin{eqnarray}
\label{eq:NoiseOperator}
{\bf \matN}_{\bf B}(t) \ge \left| i {\boldsymbol \Sigma'} - i {\bf \matD}_{\bf B}(t) \cdot {\boldsymbol \Sigma} \cdot {{\bf \matD}^T_{\bf B}(t)}  \right|,
\end{eqnarray}
where $|\cdot |$ indicates the matrix absolute value and $i{\Sigma}_{ij} \equiv \expect{[\hLambda_i,\hLambda_j]}$ and ${\boldsymbol \Sigma'}$ are the commutation matrices before and after the transformation \cite{DemoenLMP1977}.  Absent an indication that the dephasing should be extra noisy, we take the equality sign in Eq. (\ref{eq:NoiseOperator}).

In addition to the dephasing of the atomic terms described above, the covariance matrix $\Gamma$ undergoes a coherent evolution as described by Eq.~(\ref{eq:GammaEvolution}) in terms of the matrix
\begin{equation}
{\bf T}_{\bf B}(t) = \left(
\begin{array}{ccc}
\mathbbm{1}_{3\times3} & \cdot & \cdot \\
\gamma\matF(t) & \gamma|{\bf B({\bf x})}|\cal{A} & \cdot \\
\cdot & \cdot & \mathbbm{1}_{3\times3}
\end{array}
\right)
\end{equation}
where the matrix $\matA$ is described above in Eq.~(\ref{eq:matA}) and
\begin{equation}
\fl
\qquad \matF_{\bf B}(t) \equiv \left(
\begin{array}{ccc}
\cdot & F_z(t) & -F_y(t) \\
-F_z(t) & \cdot & F_x(t) \\
F_y(t) & -F_x(t) & \cdot \\
J_l (t) & J_k(t) & -2J_y(t) \\
-J_k(t) & J_l(t) & 2J_x(t) \\
J_y(t) & \sqrt{3}J_m(t)-J_x(t) & -J_l(t) \\
-\sqrt{3}J_m(t)-J_x(t) & -J_y(t) & J_k(t) \\
\sqrt{3}J_l(t) & -\sqrt{3}J_k(t) & \cdot 
\end{array}
\right).
\end{equation}

Since the terms in $\matF_{\bf B}(t)$ explicitly depend on the average atomic vector $\Lambdabar(t)$ at time $t$, we update the covariance matrix in small time steps $\tau \ll\omega_0$ and keep track of the $\Gamma(t)$ and $\Vbar(t)$ at each step.
The covariance matrix finally evolves according to
\begin{equation}
 \label{eq:GammaFieldEvolution}
{\Gamma}(t + \tau ) = {\bf D}_{\bf B} \cdot {\bf T}_{\bf B} \cdot \Gamma(t) \cdot {[{\bf T}_{\bf B}]^T} \cdot {[{\bf D}_{\bf B}]^T} + {\bf N}_{\bf B} 
\end{equation}
where ${\bf D}_{\bf B}\equiv \mathbbm{1}_{3\times3}\oplus\matD_{\bf B}\oplus\mathbbm{1}_{3\times3}$ and ${\bf N}_{\bf B}\equiv \mathbbm{O}_{3\times3}\oplus\matN_{\bf B}\oplus\mathbbm{O}_{3\times3}$.

\subsection{Combined effects}
To combine the coherent rotations with decoherence, we follow~\cite{KoschorreckJPAMOP2009} and apply alternately Eqs. (\ref{eq:vbarEvolution}) \& (\ref{eq:GammaEvolution}),  Eqs. (\ref{eq:vbardecohere})  \& (\ref{eq:GammaEvolutionOpticalDecoherence}) and Eqs. (\ref{eq:DynamicsUnderB}) \& (\ref{eq:GammaFieldEvolution}) after subdividing the time interval into sufficiently small steps $\tau$. 
Note that $\tau$ is typically much smaller during the optical pulses than in the time between them, but in both cases we continue to subdivide the time evolution until the results converge.

\section{Measurement}
\label{sec:Measurement}

After passing through the ensemble, the optical pulses are typically measured by balanced polarimetry.  
This allows one component $\hat{S}_{\rm det}\supm$ of $\Shat\supm$ to be detected.  
In the process, the pulse is absorbed by the detector and any possible information about the other components of $\Shat\supm$ is lost.   
We can describe the projective measurement by a unit vector ${\bf p}\supm$, defined such that ${\bf p}\supm \cdot \Vhat = \hat{S}_{\rm det}\supm$.

The result of the measurement is random and gaussian-distributed, with a variance $\var{\hat{S}_{\rm det}} = {\bf p}\supm \cdot \Gamma_{\bV} \cdot {\bf p}\supm$.  
The correlation of $\hat{S}_{\rm det}\supm$ with other variables are given by the vector $\Gamma_{\bV} \cdot {\bf p}\supm$.  
Finally, we can calculate the post-measurement uncertainty  in $\Vhat$, in light of the measurement result and the known correlations, to find the posterior uncertainty
\begin{eqnarray}
\Gamma_{\bV}' &=& \Gamma_{\bV} - \frac{(\Gamma_{\bV}\cdot {\bf p}\supm) \wedge (\Gamma_{\bV}\cdot {\bf p}\supm)}{{\bf p}\supm \cdot \Gamma_{\bV} \cdot {\bf p}\supm}\nonumber \\ & =&  \Gamma_{\bV} - \Gamma_{\bV} [\Pi_{{\bf p}\supm}  \Gamma_{\bV} \Pi_{{\bf p}\supm} ]^{({\rm MP})} \Gamma_{\bV}^T
\end{eqnarray}
where the superscript $^{(\rm MP)}$ indicates the Moore-Penrose pseudo-inverse and $\Pi_{{\bf p}\supm} = {\bf p}\supm \wedge {\bf p}\supm$ is a projector.

It should be noted that there is nothing stochastic in the simulation method:  only the average values, variances and covariances are computed.  
This again relies on the assumption that the fluctuations remain within a linear regime.

 \section{Initial state and technical noise contributions}
 \label{sec:TechnicalNoise}
 
A  variety of initial states can be used with this approach, subject to a few limitations:  The average value $\Vbar$ must be physical, i.e., within the range of $\Vhat$ for $\NA$ atoms and $\NL$ photons.  
The Robertson-Schr\"odinger  uncertainty principle 
\begin{eqnarray}
\delta A^2 \delta B^2 \ge \frac{1}{4} \left| \lowexpect{[A,B]} \right|^2
\end{eqnarray}
places lower limits on the uncertainty implied by $\Gamma_{\bV}$.  
Given any two scalar operators $\hat{a} \equiv {\bf a} \cdot \Vhat$, $\hat{b} \equiv {\bf b} \cdot \Vhat$
\begin{eqnarray}
\delta{a}^2 \, \delta{b}^2  =( {\bf a} \cdot \Gamma_{\bV} \cdot {\bf a} )(    {\bf b} \cdot \Gamma_{\bV} \cdot {\bf b} )  
 \ge \frac{1}{4} \left| a_i b_j \struc{\vnv_i}{\vnv_j}{\vnv_k} {\bar{\rm v}_k} \right|^2.
\end{eqnarray}

In many situations it is appropriate to assume an initial state of the form $R \equiv \rho^{\otimes \NA}$ for the atoms.  
For example, if optical pumping is used to initialize the state it is often reasonable to assume the atoms are independently pumped into a state $\rho$.  
This gives $\Lambdabar = \NA {\rm Tr}[\lambdahat \rho]$ and $\Gamma_{\Lambdab} = \NA \Gamma_{\lambdab}$.  
If $\rho$ is a pure state, then $R$ describes a coherent spin state (CSS).

We note an important subtlety about using CSSs for practical modeling:  if $\rho$ is a pure state, it is fully-polarized along some spin direction $\lambdab_{\rho}$, with no uncertainty in that direction.  
This manifests as a zero eigenvalue of $\Gamma_{\lambdab}$, and, if we na\"{\i}vely apply the above, also in $\Gamma_{\Lambdab}$.  
This is unrealistic, however.  
In practice, $\NA$ is usually determined by a physical process with significant uncertainty, e.g. trap loading.  
Indeed, $\delta \NA^2$ is often larger than the quantum variances $\NA \Gamma_{\lambdab} \sim {\NA}$.  
This contributes a significant ``atomic technical noise'' to $\Gamma_{\Lambdab}$.  
As shown in  \ref{app:NANoise}, for a state of the form $R=\rho^{\otimes \NA}$, where $\NA$ has average $\NAb$ and  variance $\delta \NA^2$,
\begin{eqnarray}
\label{eq:TechnicalNoise}
\Gamma_{\bf \Lambda}  & = & \NAb \Gamma_\lambdab + \delta\NA^2 (  \bar{\boldsymbol \lambda} \wedge  \bar{ \boldsymbol \lambda}).
\end{eqnarray}

In most situations of interest the input light state is a coherent state. 
For example, in the experiments described below, we use an $\hat{S}_x$-polarized input, i.e. $\Sbar=(\NL/2,0,0)$, which has a covariance matrix $\Gamma_{\bf S}={\rm diag}(\NL/4, \NL/4, \NL/4)$.
The input average magnetic field components $\Bbar$ and covariance matrix $\Gamma_{\bf B}$ will depend on the modeled experiment.

In the experiments described below the atoms and light are independently initialized, and we assume that the atom, light and magnetic field variables are initially uncorrelated.
Thus the initial covariance matrix of eq. (\ref{GammaV}) can be written:
 \begin{equation}
\Gamma_{\bf V}  = \Gamma_{\bf B}\oplus \Gamma_{\bf \Lambda}\oplus\Gamma_{\bf S}.
\label{GammaVInit}
\end{equation}

\section{An example: Free-induction decay of collective atomic spin}
\label{sec:FID}

We illustrate the formalism described above with a simple example.
We study, experimentally and theoretically, a paramagnetic Faraday rotation measurement of the free--induction decay (FID) of input $\hat{F}_z$ and $\hat{F}_y$-polarized coherent spin states (CSS) precessing in a magnetic field.
As described in Ref.~\cite{BehboodAPL2013}, neglecting tensorial light shifts described by the $G_2$ term of Eq.~\ref{Heff}, this leads to a measurable ploarization rotation angle
\begin{equation}
\fl \ave{\phi(t)} = \frac{G_1}{|B|^2} \times \cases{
	\left[ B_y B_z \left(1-\cos(\gamma |B| t)  e^{-{t}/{T}}\right) + B_x |B|\sin(\gamma B t)  e^{-{t}/{T}} \right] \ave{\hat{F}_y(0)} & \\
	\left[{B_z^2}+[{B_x^2+B_y^2}]\cos(\gamma |B| t) e^{-{t}/{T}} \right] \ave{\hat{F}_z(0)}\\
	}
\label{eq:FIDSignal}
\end{equation}
for the input $\hat{F}_z$ and $\hat{F}_y$-polarized CSS respectively.
The transverse relaxation time $T= 1/(\width\gamma |B'_\parallel |)$ is due to the field-parallel gradient component $B'_\parallel \equiv \partial |{B}| /\partial_z$, and a Lorentzian distribution (full-width at half-maximum $\width$) of atoms along $z$, the trap axis, as described in \ref{app:inhomogeneous}.
Tensorial light shifts induce an additional nonlinear rotation of the atomic spins, as described in Refs~\cite{Deutsch2010,Smith2004}.
This measurement can be used to estimate an unknown vector magnetic field, as in Ref.~\cite{BehboodAPL2013}, and is the basis of a proposal to prepare a planar--squeezed atomic spin state state~\cite{Puentes2012}.

\begin{figure}
\centering
\includegraphics[width=\textwidth]{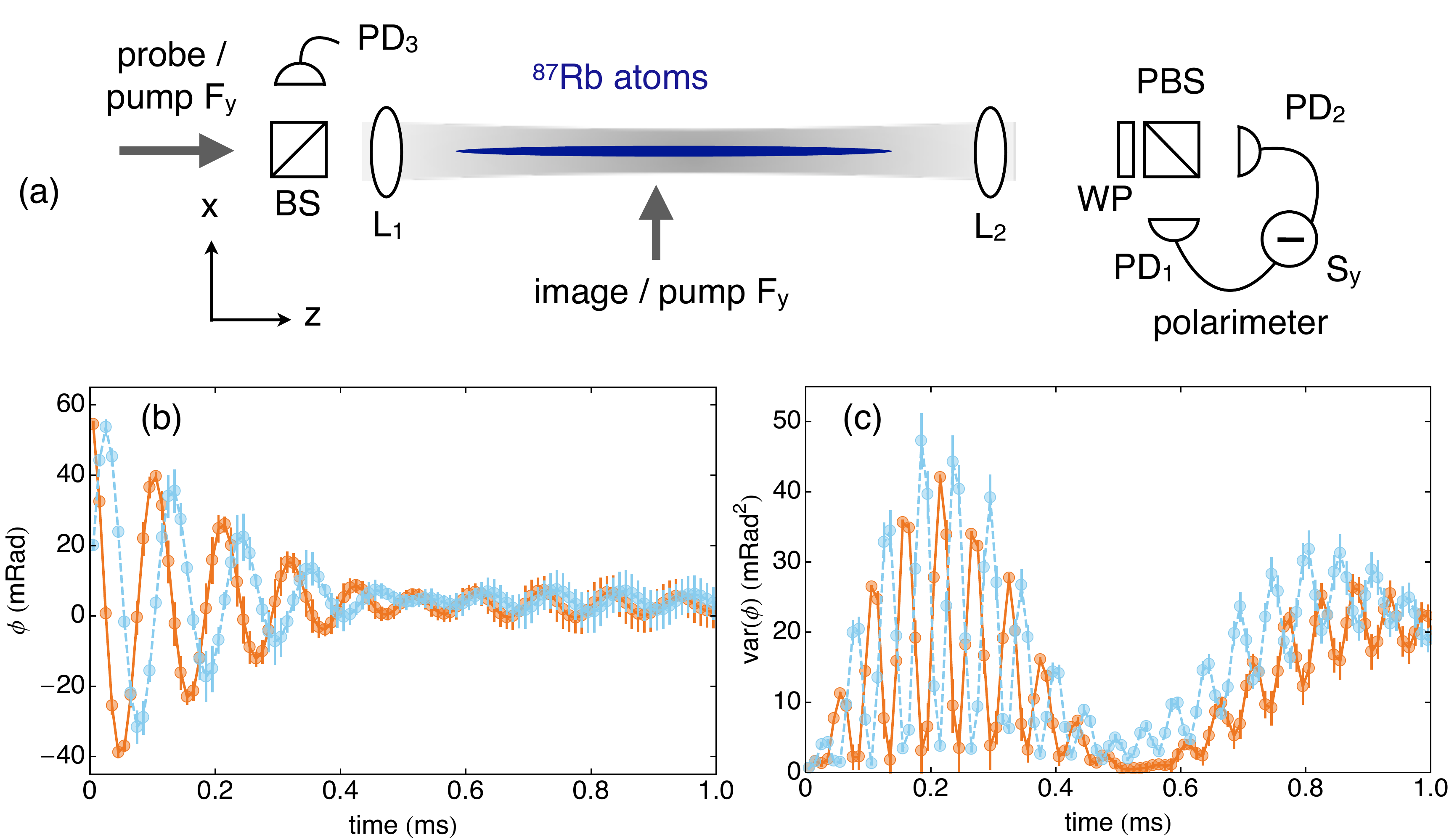}
\caption{
(a) Experimental setup.  PD: photodiode; L: lens; WP: wave plate; BS: beam splitter; PBS: polarizing beam splitter.
Atoms are loaded into a single--beam optical dipole trap and probed with pulses of light propagating along the trap axis.
An external magnetic field is applied to coherently rotate the atomic spins, and gradient field components $\partial B_i/\partial z$ are actively cancelled.
The initial atomic state is prepared via optical pumping with circularly polarized light {either} propagating perpendicular to or along the trap axis, to prepare an $\hat{F}_y$-polarized or $\hat{F}_z$-polarized CSS, respectively.
Also shown are experimental data of (b) the average signal $\bar{\phi}$ for an input $\hat{F}_y$ -- and $\hat{F}_z$--polarized CSS (blue and orange respectively), and (c) the evolution of the variance $\var{\phi}$ for the same input states. 
Error bars represent $\pm1\sigma$ statistical errors.
\label{fig:magnetometer}}
\end{figure}

Our experimental apparatus, illustrated in Fig.~\ref{fig:magnetometer}, has been described in detail elsewhere~\cite{SewellPRL2012,KoschorreckPRL2010b,KoschorreckPRL2010a,KubasikPRA2009}.
In brief, we work with an ensemble of up to one million $^{87}$Rb atoms held in a weakly focused single beam optical dipole trap and probed by \unit{\micro\second} duration pulses of near--resonant light propagating along the trap axis and focused to a spot matching the radial width of the atomic cloud.
This geometry produces a strong atom--light coupling, characterized by the effective on-resonance optical depth $d_0\equiv(\sigma_0/A)\NA$, where $\sigma_0$ is the effective on--resonance scattering cross--section, $A$ is an effective atom--light interaction area and $\NA$ is the number of atoms~\cite{KubasikPRA2009}.
In this experiment we observe an effective optical depth $d_0=29.0(2)$.
The optical readout can achieve projection-noise-limited sensitivity, calibrated against a thermal spin state~\cite{KoschorreckPRL2010a}, and using dynamical decoupling techniques~\cite{KoschorreckPRL2010b} has been used to demonstrate spin squeezing and entanglement--enhanced measurement sensitivity of the collective atomic spin~\cite{SewellPRL2012}.

For the FID measurement the atoms are initially polarized via optical pumping with a \unit{5}{\micro\second} duration pulse of circularly polarized light tuned to the $f=1\rightarrow f'=1$ transition of the ${\rm D}_2$ line.
The optical pumping pulse propagates either along the trap axis to produce an $F_z$-polarized CSS ($\ave{\hat{F}_x(0)}=\NA$), or perpendicular to the trap to produce an $F_y$-polarized CSS ($\ave{\hat{F}_y(0)}=\NA$).
Simultaneously with the optical pumping pulse, light tuned to the $f=2\rightarrow f'=2$ transition of the ${\rm D}_2$ line is applied to prevent atoms collecting in the $f=2$ hyperfine state.
This ensures that we prepare the initial CSS with good fidelity.
We estimate an optical pumping efficiency of $>99\%$ for both the $\hat{F}_z$--polarized and $\hat{F}_y$-polarized CSSs.
The remaining unpolarized atoms make a negligible contribution to the observed $\var{\phi}$.

\begin{figure}[ht]
\centering
\includegraphics[width=\textwidth]{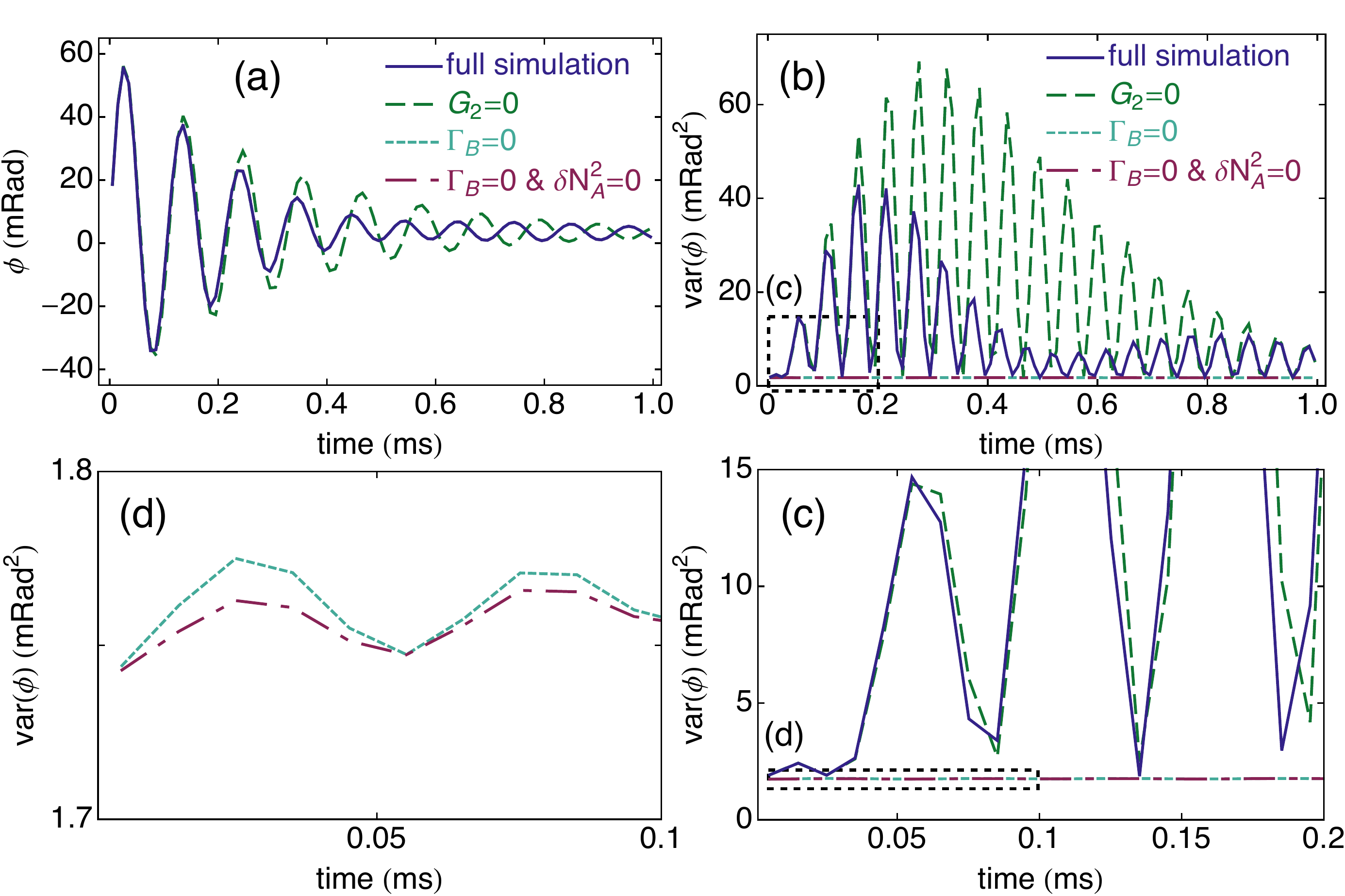}
\caption{Illustration of the effect of technical noise terms and tensorial light shifts on the mean $\bar{\phi}$ and variance $\var{\phi}$ of the observed signal.
(a) Tensorial light shifts rotate population out of $\hat{F}_z$ and into the alignment variables $\hat{J}_{x,y}$ leading to a collapse and revival of the mean $\bar{\phi}$.
The revival is accompanied by a $\pi$ phase shift in the oscillations, as is evident in the simulation with $G_2=0$, which removes the effect of tensorial light shifts.
(b)--(d) The observed variance $\var{\phi}$ undergoes a more pronounced collapse and revival driven by the tensorial light shifts, which couple technical noise from the atomic variables $\Lambdahat$ and the magnetic field covariance matrix $\Gamma_{\bf B}$ into the observed variable.
The contribution to the observed signal in (b) is due to magnetic field noise.
Technical noise in the atomic variables due to uncertainty $\delta\NA^2$ in the atom number only adds significant noise only during the early stages of the evolution.
This is illustrated more clearly in the magnified plots (c) and (d).
The dot-dashed magenta line in plot (d) illustrates the quantum noise contributions due to the light (shot--noise) and atoms (projection--noise).
\label{fig:TechnicalNoise}}
\end{figure}
The atoms are then probed with a sequence of \unit{1}{\micro\second} long pulses of linearly polarized light ($\ave{\hat{S}_x^{\rm (in)}}=\pm\NL/2$) detuned about \unit{1}{\giga\hertz} from resonance with the $f=1\rightarrow f'=0$ transition on the ${\rm D}_2$ line and propagating along the trap axis.
After each pulse, the Faraday rotation signal $\phi=\hat{S}_y^{\rm (out)}/\hat{S}_x^{\rm (in)}$ is recorded  by a shot--noise limited polarimeter~\cite{WindpassingerMST2009}.
The pulses are either (a) all h--polarized and sent through the atoms at $P_1=\unit{10}{\micro\second}$ intervals, or (b) pairs of alternately h-- and v--polarized pulses separated by \unit{3}{\micro\second} and sent through the atoms $P_2=\unit{20}{\micro\second}$ intervals.
The entire experimental sequence is repeated up to 400 times to collect statistics.

Typical experimental data  are shown in Fig.~\ref{fig:magnetometer}; we plot the evolution of the average signal $\bar{\phi}$ for an input $\hat{F}_y$-- and $\hat{F}_z$--polarized CSS , and the evolution of the variance $\var{\phi}$ for the same input states.
The observed signal initially oscillates at the Larmor precession frequency $\omega_0=2 \pi \times 9.2$~kHz with an exponential decay, as described by Eq.~(\ref{eq:FIDSignal}).
On top of this there is a collapse and revival of the oscillations driven by tensorial light shifts, as described in Refs.~\cite{Deutsch2010,Smith2004}, which rotate population out of $\hat{F}_z$ and into the alignment variables $\hat{J}_{x,y}$ at a rate $\omega_{G_2}=(G_2 S_x/2)(1/P_1) =2 \pi \times 0.43$~kHz.
The revival is accompanied by a $\pi$ phase shift in the oscillations (see Fig.~\ref{fig:TechnicalNoise} below for an illustration of this effect).
The observed variance $\var{\phi}$ oscillates at a frequency $2\omega_0$ and undergoes a similar but much more pronounced collapse and revival driven by the tensorial light shifts.
Note that the observed variance in this experiment is dominated by technical noise; for our experimental parameters, the quantum noise contributions from both the atomic and light variables are $\ll \unit{1}{\milli\rad^2}$.
The dominant technical noise contributions are due to uncertainty in the atom number $\delta\NA^2$, and magnetic field noise, described by the covariance matrix $\Gamma_{B}$, coupled into the observed variable via tensorial light shifts.
We illustrate the effect of these terms separately in Fig.~\ref{fig:TechnicalNoise}, by running numerical simulations, as described in detail in Section~\ref{sec:Simulation}, using experimental parameters from the data shown in Fig.~\ref{fig:magnetometer} and setting variously the $\Gamma_{\bf B}$, $\delta N^2$ and $G_2$ terms equal to zero.

\section{Simulation of free--induction decay}
\label{sec:Simulation}

In order to simulate the FID experiment, we need to estimate a number of experimental parameters, including the input state vector $\Vbar$ and covariance matrix $\Gamma_{\bf V}$, the coupling constants of Eq.~(\ref{Heff}), and the decoherence terms in Eqs.~(\ref{DL})--(\ref{NS}) and Eq.~(\ref{eq:ExponentialDephasing}).
The atom--light coupling constant $G_{1}$  is calibrated in an auxiliary experiment as described in Ref.~\cite{KubasikPRA2009}.
From this calibration we calculate the effective atom--light interaction area $A$, the coupling constant $G_{2}$ of Eq.~(\ref{Heff}) and the single--photon scattering probability $\eta_{\gamma}$ used to describe the optically induced decoherence terms in Eqs.~(\ref{DL})--(\ref{NS}) (see Ref.~\cite{KoschorreckT2010} for details).

We estimate the average magnetic field $\bar{\bf B}$ for a single experimental data set by fitting the observed signal $\ave{\phi(t)}$ with Eq.~(\ref{eq:FIDSignal}), as described detail in Ref.~\cite{BehboodAPL2013}.
Since Eq.~(\ref{eq:FIDSignal}) neglects the effect of tensorial light shifts, we fit only the first \unit{250}{\micro\second} of each measured signal to minimize the systematic error that this approximation introduces.
From the fits we also determine the coherence time $T$ that is used in Eq.~(\ref{eq:ExponentialDephasing})
From the same fits we estimate the number of atoms $\ave{\hat{F}_i(0)}= \NA/2$.
Assuming an input state of the form $R \equiv \rho_{\hat{f}_i}^{\otimes \NA}$ for the atoms, this specifies the inititial atomic state via $\Lambdabar = \NA {\rm Tr}[\lambdahat \rho_{\hat{f}_i}]$.
The number of photons $\NL$ in each pulse is independently measured via a reference detector, as shown in Fig.~\ref{fig:magnetometer}.
Together, these estimates specify the initial state of the vector $\Vbar$.

Statistics from the fits to Eq.~(\ref{eq:FIDSignal}) across the data set allow us to estimate the covariance matrix $\Gamma_{\bf B}$.
Similarly, we estimate the uncertainty $\delta \NA^2$ in the atom number from $\var{\hat{F}_i}=\var{\NA/2}$ (which includes contributions from variable trap loading and state preparation efficiency as well as measurement uncertainty).
This allows us to estimate the initial covariance matrix $\Gamma_{\mathbf{\Lambda}}$ using Eq.~(\ref{eq:TechnicalNoise}).
Since the measurement is shot--noise limited, the input light covariance matrix is $\Gamma_{\bf S}=(\NL/4,\NL/4,\NL/4)$.
Together, via Eq.~(\ref{GammaVInit}), these specify the initial covariance matrix $\Gamma_{\bf V}(0)$.

As an example, we give the experimental parameters in detail of the first example described in Section~\ref{sec:FID}, shown in Fig.~\ref{fig:magnetometer}(b) and~(c).
For this experiment, we set the detuning of the probe to $\Delta=-\unit{700}{\mega\hertz}$ and probe with a sequence of \unit{1}{\micro\second} long h--polarized pulses of light with on average $\NL=7.2\times10^6$ photons per pulse at \unit{10}{\micro\second} intervals.
The measured average magnetic field was ${\bf{B}}=\unit{(11.98,-4.38,-4.01)}{\milli G}$, with a covariance matrix
\begin{equation}
\label{eq:GammaBExpt}
\Gamma_{\bf B} = \left(
\begin{array}{ccc}
 0.202& 0.0373& -0.048\\
 0.037& 0.201& 0.016\\
 -0.048& 0.016& 0.019
\end{array}
\right){\rm mG^2}.
\end{equation}
We estimate $\NA=61.7\times10^5$ and $\Delta \NA = 1.4\times10^3$, giving an initial atomic vector for the $\hat{F}_y$--polarized input ${\bf \Lambda}(t=0)=(0, 1, 0, -0.5, 0, 0, 0, -0.29)\times\NA$ and covariance matrix
\begin{equation}
\fl
\qquad
\Gamma_{\bf \Lambda}(t=0)= \left(
\begin{array}{cccccccc}
 0.5 & 0 & 0 & 0 & 0.5 & 0 & 0 & 0 \\
 0 & 11.15 & 0 & -5.58 & 0 & 0 & 0 & -3.22 \\
 0 & 0 & 0.5 & 0 & 0 & 0 & 0.5 & 0 \\
 0 & -5.58 & 0 & 3.04 & 0 & 0 & 0 & 1.18 \\
 0.5 & 0 & 0 & 0 & 0.5 & 0 & 0 & 0 \\
 0 & 0 & 0 & 0 & 0 & 1 & 0 & 0 \\
 0 & 0 & 0.5 & 0 & 0 & 0 & 0.5 & 0 \\
 0 & -3.22 & 0 & 1.18 & 0 & 0 & 0 & 1.68
\end{array}
\right)\times \NA.
\end{equation}
We further estimate a coherence time $T=\unit{360\pm10}{\micro\second}$.
The calibrated coupling constants were $G_1=1.7\pm0.2\times10^{-7}$ radians per atom and $G_2=-7.5\pm0.8\times10^{-9}$ radians per atom, and the atom--light scattering parameter $\eta_{\gamma}=1.1\times 10^{-9}$.

\begin{figure}
\centering
\includegraphics[width=\textwidth]{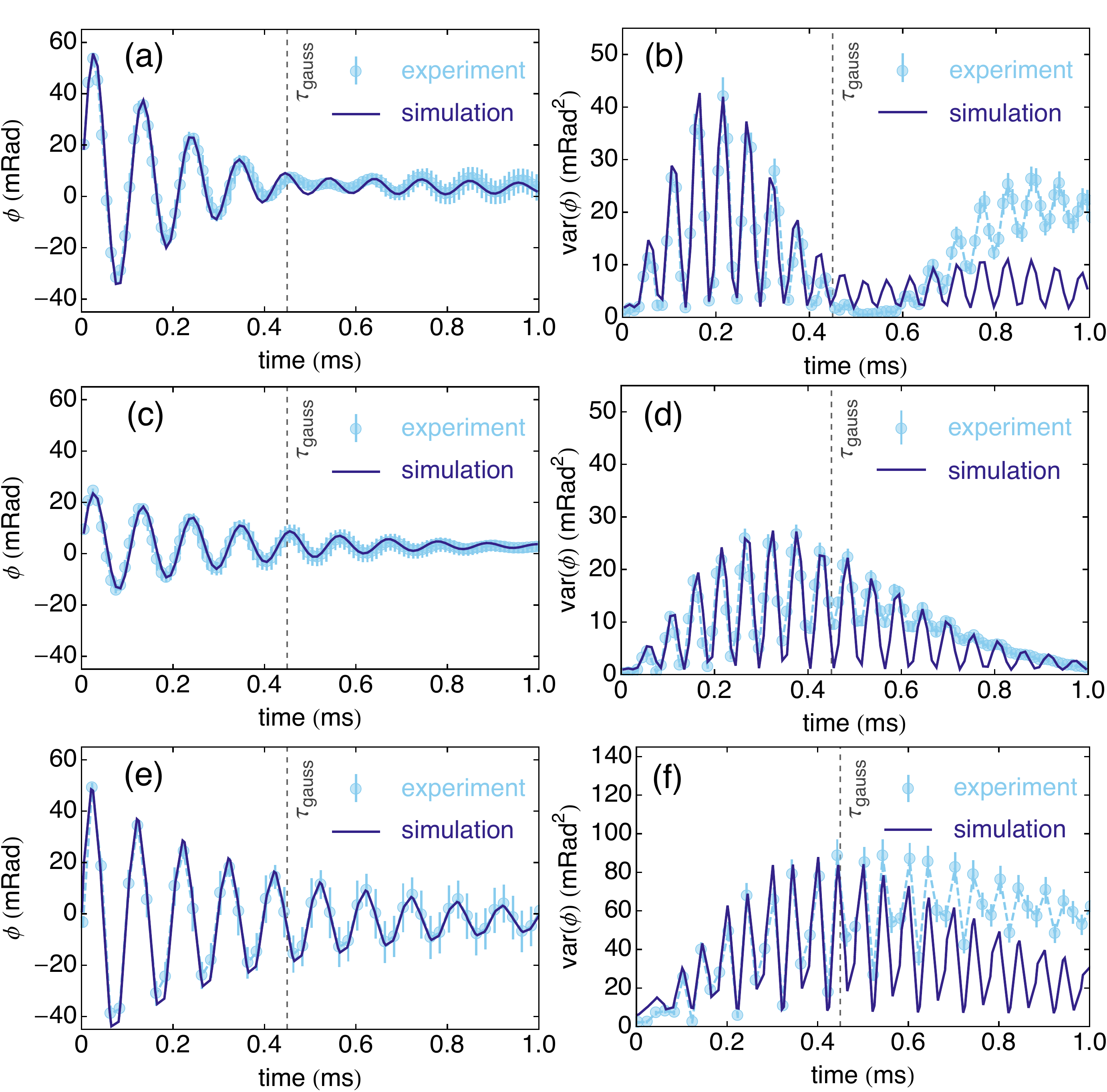}
\caption{
Comparison of experimental data (light blue circles) with numerical simulations (dark blue curves) of the mean $\bar{\phi}$ (a), (c) and (e) and variance $\var{\phi}$ (b), (d) and (f) of the optical rotation of the probing light.
For (a) and (b) we set the detuning of the probe to $\Delta=-\unit{700}{\mega\hertz}$ and probed the atoms with a sequence of \unit{1}{\micro\second} long h--polarized pulses of light with on average $\NL=7.2\times10^6$ photons per pulse at \unit{10}{\micro\second} intervals.
For (c) and (d) we set the detuning of the probe to $\Delta=-\unit{1.5}{\giga\hertz}$ and probed with the same measurement sequence. 
For (e) and (f) we set the detuning of the probe to $\Delta=-\unit{700}{\mega\hertz}$ and probed the atoms with a sequence pairs of \unit{1}{\micro\second} long pulses with alternating h-- and v--polarization, separated by \unit{3}{\micro\second} and sent through the atoms \unit{20}{\micro\second} intervals.
Error bars represent $\pm1\sigma$ statistical errors.
See text for details.
\label{fig:ExpTheory}}
\end{figure}

Together these parameters determine the initial state vector $\Vbar(t=0)$ and covariance matrix $\Gamma_{\bf V}(t=0)$ as described in Section~\ref{sec:TechnicalNoise}, the atom--light coupling constants of Eq.~(\ref{Heff}), the magnetic field components of Eq.~(\ref{eq:dynamicalFieldEquations}), and the decoherence terms in Eqs.~(\ref{DL})--(\ref{NS}) and Eq.~(\ref{eq:ExponentialDephasing}). 
With these input parameters, we then run numerical simulations of the evolution of the state vector $\Vbar$ and covariance matrix $\Gamma_{\bf V}$ following the procedure described in Section~\ref{sec:Dynamics}, keeping track of ${\bf V}(t_m)$ and $\Gamma_{\bf V}(t_m)$ at each time step.
As in the experiment, we alternate between a time interval of $\unit{1}{\micro\second}$ in which the light is present and an interval of $\unit{9}{\micro\second}$ with no light present.
For the numerical calculations, we divide these intervals into 50 and 100 sub-steps respectively, which is sufficient to ensure numerical convergence of the results.
The results of the simulations (dark blue curves) are plotted along with the experimental data (light blue circles) in Fig.~\ref{fig:ExpTheory}(a) and~(b) for both the mean $\bar{\phi}$ and variance $\var{\phi}$ with an input $\hat{F}_y$--polarized atomic state.

We observe excellent qualitative and quantitative agreement between the simulations and the observed mean $\bar{\phi}$ and variance $\var{\phi}$ in the rotation angle with no further free parameters adjusted in the calculations.
For the mean $\bar{\phi}$, we reproduce the observed behaviour in our simulations over the entire $\unit{1}{\milli\second}$ of observation, as shown in Figs.~\ref{fig:ExpTheory}(b) and~(d).
For the variance $\var{\phi}$ the quantitative agreement is initially good, but breaks down at longer times.
This can be explained by the effect of the uncertainty in the magnetic field, described by the covariance matrix $\Gamma_{\bf B}$, which eventually drives the atomic variables out of the gaussian approximation.

Because magnetic precession is cyclic, not linear, an initially gaussian spin distribution will become non-gaussian due to uncertainty in the precession frequency.  As some parts of the distribution precess faster than others, the distribution begins to ``wrap around'' the Bloch sphere, forcing a non-gaussian shape on the distribution.  It is convenient to define  a time $\tgauss\equiv\pi/(\gamma_F\Delta B_{\parallel})$, where $\Delta B_{\parallel}$ is the uncertainty in ${\bf B}$ along the average field direction, found using $\Gamma_{\bf B}$ of Eq. (\ref{eq:GammaBExpt}).  $\tgauss$ indicates the moment at which the precession angle uncertainty becomes $\pi$.   Perhaps surprisingly, the observed and predicted variances (see Fig.\ref{fig:magnetometer}(c) and~(e)) agree very well up to $t=\tgauss \approx 0.45~{\milli\second}$, showing that the theory gives quantitatively accurate results even for significant departures from gaussianity.  The agreement in the {\it average} values persists even for $t>\tgauss$. 
This changes accumulates with time and becomes much more significant than the technical noise in the number of atoms so that while the atomic noise is important early on, the dephasing effect of magnetic field inhomogeneities becomes the dominant contribution for larger times.

The collapse and revival of the oscillations in Figs.~\ref{fig:ExpTheory}(a) and~(b) is due to rotations driven by the tensorial light shift.
The effect of the tensorial light shifts can be reduced either by probing further off resonance, or by probing the atoms alternately with h-- and v--polarized pulses, as described in detail in Ref.~\cite{Koschorrek2010}.
We illustrate this in Fig.~\ref{fig:ExpTheory}(c)--(d), where we compare these data to FID measurements (and simulations) made with two alternate probing strategies.
In Fig.~\ref{fig:ExpTheory}(c) and (d) we set the detuning to $\Delta=\unit{-1.5}{\giga\hertz}$ and repeat the single--polarization probing sequence.
In Fig.~\ref{fig:ExpTheory}(e) and (f) we set the detuning to $\Delta=\unit{-700}{\mega\hertz}$ and probe with pairs of alternately h-- and v--polarized pulses separated by \unit{3}{\micro\second} are sent through the atoms \unit{20}{\micro\second} intervals (for clarity we plot only the h--polarized pulses).
This results in the same total number of photons used per unit time as in the single polarization probing strategy.
With these data we observe similar behavior in the mean $\bar{\phi}$ and variance $\var{\phi}$ at both detunings, indicating the effective cancellation of tensorial light shifts in these experiments.
We also observe non--zero minima in $\var{\phi}$ at all phases in the experimental data for $t>\unit{0.5}{\milli\second}$, which are not reproduced in our calculations.
This may due to the finite sampling time of the measured data, which are not included within our calculations or in terms of the curvature of the Bloch sphere: for small uncertainty in precession angle $\Delta \theta = t \gamma_F \Delta B_\parallel$, we will have $\Delta F_z \propto \Delta \theta \partial_\theta F_z$.  For flat parts of the curve $F_z(\theta)$ there will be zero $\partial_\theta F_z$ and thus zero $\Delta F_z$  but for larger $\Delta \theta$, we need to take into account higher derivatives, i.e. the curvature $\partial_\theta^2 F_z$ and higher.
Finally we note that our model works always well for $t<\tgauss$ where we expect the Gaussian approximation to hold, as described above. 

\section{Conclusions}
\label{sec:Conclusion}

We have presented a method for describing the quantum dynamics of spin-1 atomic ensembles, extending the method introduced for spin-1/2 atomic ensembles by Madsen and M{\o}lmer \cite{MadsenPRA2004} and generalized by Koschorreck et al. \cite{KoschorreckJP2009} and Toth et al.  \cite{TothJP2010}.  
Our approach, which explicitly includes the so-called ``spin alignment'' degrees of freedom, fits naturally with the light-matter interaction, which couples both spin alignment and spin orientation to the optical Stokes parameters.  
For spin-1 our description is complete within the gaussian approximation, while for larger spins  it is still useful when octopole and higher spin moments can be neglected.  
We also include the important technical noise associated with magnetic fields and noise due to uncertain atom number, as typically arises due to stochastic trap-loading processes.  
Finally, we give explicit formulae for the noise introduced by spontaneous scattering during the optical probing process and due to dephasing in an inhomogeneous magnetic field.  

We have tested the model against experiment in a scenario involving all of these effects.  
We compute the evolution of  the spin orientation average and variance for atomic ensembles with uncertain atom number, undergoing a combination of free-induction decay and alignment-to-orientation conversion in the presence of a noisy magnetic field.  
The simulation is compared to experimental observations made with a cold $^{87}$Rb ensemble held in an optical dipole trap, probed by shot-noise-limited Faraday rotation with near-resonant light.  
We find good agreement within the gaussian regime.  
In addition to validating the model, the experiments provide a heuristic guide to the limits of the gaussian approximation in these systems. 

Given that most atomic ensemble experiments are performed with spin-1 or larger atoms, the technique described here will allow more accurate modeling of established quantum optical protocols, e.g. quantum memory \cite{HammererPR2004}, quantum non-demolition measurement \cite{SewellARX2013,KoschorreckPRL2010a}, dynamical decoupling \cite{KoschorreckPRL2010b},  spin squeezing \cite{SewellPRL2012} and vector magnetometry \cite{BehboodAPL2013}, as well as proposed applications such as generation of macroscopic singlet states \cite{TothJP2010,UrizarARX2012} and planar squeezed states \cite{Puentes2012}.

\section{Acknowledgements}
This work was supported by the Spanish MINECO under the project MAGO (ref. no. FIS2011-
23520), by the European Research Council under the project AQUMET, and by Fundaci\'{o} Privada Cellex.

\bigskip
\section*{References}
\bibliographystyle{unsrt}

\appendix

\section{Proof of Eq.~(\ref{eq:GammaFromDecoherenceAgain})}
\label{app:decoherence}
For brevity, we write the covariance as
\begin{eqnarray}
 \cm{\bf A, B} & \equiv & \frac{1}{2} \ave{{\bf A} \wedge {\bf B}+({\bf B} \wedge {\bf A})} - \ave{{\bf A}} \wedge \ave{{\bf B}} 
\end{eqnarray}
where the expectation is taken on the state of the ensemble.  
In this notation, $\Gamma_{\Lambdab}^{} \equiv \cm{\Lambdahat,\Lambdahat} $ where as above $\Lambdahat = \sum_i \lambdahat^{(i)}$.  
We assume that all atoms are statistically equivalent, so that  $\ave{ \lambdahat^{(i)} \wedge \lambdahat^{(j)} } = \ave{ \lambdahat^{(1)} \wedge \lambdahat^{(2)} }$ for all $i\ne j$ and $\ave{ \lambdahat^{(i)} \wedge \lambdahat^{(i)} } = \ave{ \lambdahat^{(1)} \wedge\lambdahat^{(1)} }$ for all $i$.  
We then have  
\begin{eqnarray}
\label{eq:GammaFromC}
\Gamma_{\Lambdab}^{} &=& \sum_{i,j} \cm{\lambdahat^{(i)},\lambdahat^{(j)} }  =  \NA \cm{\lambdahat^{(1)},\lambdahat^{(1)} }  + \NA(\NA-1) \cm{\lambdahat^{(1)},\lambdahat^{(2)}} \nonumber \\
\end{eqnarray}
which we solve for $ \cm{\lambdahat^{(1)},\lambdahat^{(2)}}$ to get 
\begin{eqnarray}
 \cm{\lambdahat^{(1)},\lambdahat^{(2)}}  & = &\frac{ \Gamma_{\Lambdab}^{} -  \NA  \cm{\lambdahat^{(1)},\lambdahat^{(1)}}}{\NA(\NA-1) }.
 \end{eqnarray}
Because of the symmetry, removing atoms does not change $ \cm{\lambdahat^{(1)},\lambdahat^{(1)}} $ or $ \cm{\lambdahat^{(1)},\lambdahat^{(2)}} $.   
If a fraction $1-\X$ of the atoms is removed, the covariance matrix $\Gamma_\Lambdab^{(\X\NA)}$ of the remaining atoms can be calculated as in Eq.~(\ref{eq:GammaFromC}), but summing $i$ and $j$ from $1$ to $\X \NA$.  
We find 
\begin{eqnarray}
\Gamma_{\Lambdab}^{(\X \NA)} &= & \X \NA \cm{\lambdahat^{(1)},\lambdahat^{(1)}}   + \X \NA(\X \NA-1) \cm{\lambdahat^{(1)},\lambdahat^{(2)}} \nonumber \\
& = & \Gamma_\Lambdab^{(\NA)} \frac{\X (\X \NA -1)}{\NA-1}
+  \Gamma_\lambdab  \left(\frac{\X (1-\X) \NA^2}{\NA-1} \right)\nonumber
\end{eqnarray}
where 
\begin{eqnarray}
\label{eq:GammalambdaDef}
\Gamma_\lambdab \equiv \cm{\lambdahat^{(1)},\lambdahat^{(1)}}
\end{eqnarray}
is the single-atom covariance matrix.
Dropping terms of order $1/\NA$ and smaller, 
\begin{eqnarray}
\label{eq:GammaFromRemoval}
\Gamma_{\Lambdab}^{(\X \NA)}
 &= & \Gamma_{\Lambdab}^{(\NA)} \X^2 +  \Gamma_\lambdab  \X (1-\X) \NA.
 \end{eqnarray}
This accounts for the change in $\Gamma_{\Lambdab}$ due to {removing} $(1-\X) \NA$ atoms.
We must also account for the noise of returning a fraction $p$ of these atoms to the ensemble in a decohered state.  
We assume they are completely random, and thus add the noise of a thermal state (i.e. variance $f(f+1)/3$ per atom).
\begin{eqnarray}
\Gamma_{\Lambdab}^{(\X \NA)}
 &= & \Gamma_{\Lambdab}^{(\NA)} \X^2 +  \Gamma_\lambdab  \X (1-\X) \NA 
+ 2 \p (1-\X)  \NA \mathbbm{1}/{3}.
\end{eqnarray}
Here 
\begin{eqnarray}
\label{eq:singleAtomsCovariance}
\Gamma_\lambdab &=&  \frac{2}{3} \mathbbm{1}_{8\times 8} - \bar{\lambdab} \wedge \bar{\lambdab} +  \sum_k \bar{\lambda}_k {\bf M}^{(k)}, \nonumber \\
\label{eq:singleAtomsCovarianveMatrices}
M^{(k)}_{ij} &\equiv&  \frac{1}{4} {\rm Tr}[ \hat{\lambda}_k \{ \hat{\lambda}_i ,  \hat{\lambda}_j \} ]
\end{eqnarray}
where $\{\cdot,\cdot\}$ indicates the anti-commutator. 
This can be shown using the expectation ${\rm Tr}[\rho \hat{\lambda}_i] = \bar{\lambda}_i$ and the orthonormality condition ${\rm Tr}[\lambda_i \lambda_j] = 2 \delta_{ij}$ from which ${\rm Tr}[\{\lambda_i,\lambda_j\}] = 4 \delta_{ij}$.  We find the single-particle state $\rho =  \frac{1}{3}\mathbbm{1} +  \frac{1}{2} \sum_i \bar{\lambda}_i \hat{\lambda}_i$ and the covariances
\begin{eqnarray}
\cov(\lambda_i,\lambda_j) &\equiv&\frac{1}{2} \lowexpect{  \{ \hat{\lambda}_i ,  \hat{\lambda}_j \} } - \lowexpect{   \hat{\lambda}_i} \lowexpect{ \hat{\lambda}_j} 
\nonumber \\
&=&  \frac{2}{3} \delta_{i,j} +  \frac{1}{4} \sum_k  \bar{\lambda}_k  {\rm Tr}[ \hat{\lambda}_k \{ \hat{\lambda}_i ,  \hat{\lambda}_j \} ]
- \bar{\lambda}_i \bar{\lambda}_j  .
\end{eqnarray}

\section{Noise from uncertain atom number}
\label{app:NANoise}
We consider the statistics of $\Lambdahat$ for ensembles with $\NA$ atoms in a permutationally-invariant product  state $R^{(\NA)} = \rho^{\otimes \NA}$, and taking a statistical average over $\NA$.  We indicate averages with subscripts, e.g. $\lowexpect{\Lambdahat}_R = \Tr[\Lambdahat R]$ indicates an expectation with respect to the state $R$, while $\lowexpect{\Lambdahat}_{R,\NA} = \lowexpect{ \Tr[\Lambdahat R^{(\NA)}]}_{\NA}$ indicates a statistical average of $\lowexpect{\Lambdahat}_R$ over the distribution of $\NA$ . Due to the structure of $R^{(\NA)}$, $\lowexpect{\lambdahat^{(k)}}_{R^{(\NA)}} =  \lowexpect{\lambdahat^{(1)}}_\rho \equiv \lowexpect{\lambdahat}_\rho$ is independent of both $k$ and of $\NA$, so that 
\begin{eqnarray}
\lowexpect{\hat{\Lambda}_i \hat{\Lambda}_j}_R &=& \sum_{k,l = 1}^{\NA} \lowexpect{\hat{\lambda}_i^{(k)} \hat{\lambda}_j^{(l)}}_\rho  \nonumber \\
& = & {\NA} \lowexpect{\hat{\lambda}_i \hat{\lambda}_j}_\rho + \NA(\NA-1)  \lowexpect{\hat{\lambda}_i}_\rho \lowexpect{\hat{\lambda}_j}_\rho
\end{eqnarray}
and thus 
\begin{eqnarray}
\lowexpect{\hat{\Lambda}_i \hat{\Lambda}_j}_{R,\NA} &=& 
{\NAb} \lowexpect{\hat{\lambda}_i \hat{\lambda}_j}_\rho + (\delta \NA^2 + \NAb^2 - \NAb)  \lowexpect{\hat{\lambda}_i}_\rho \lowexpect{\hat{\lambda}_j}_\rho \nonumber \\
\end{eqnarray}
where $\delta \NA^2 \equiv \lowexpect{\NA^2} - \lowexpect{\NA}^2$ indicates the variance.  At the same time
\begin{eqnarray}
\lowexpect{\hat{\Lambda}_i}_{R,\NA} &=& 
{\NAb} \lowexpect{\hat{\lambda}_i}_\rho 
\end{eqnarray}
so that
\begin{eqnarray}
\cov(\hat{\Lambda}_i ,\hat{\Lambda}_j)_{R,\NA} &\equiv& \frac{1}{2} \lowexpect{\hat{\Lambda}_i \hat{\Lambda}_j+\hat{\Lambda}_j \hat{\Lambda}_i}_{R,\NA} - \lowexpect{\hat{\Lambda}_i}_{R,\NA} \lowexpect{\hat{\Lambda}_j}_{R,\NA} 
\\ & = & \NAb \cov(\hat{\lambda}_i, \hat{\lambda}_j)_\rho + \delta\NA^2  \lowexpect{\hat{\lambda}_i}_\rho \lowexpect{\hat{\lambda}_j}_\rho
\end{eqnarray}
In terms of the single-atom covariance matrix $\Gamma_{\lambdab}$ of Eq.~(\ref{eq:GammalambdaDef}),
\begin{eqnarray}
\Gamma  & = & \NAb \Gamma_{\lambdab} + \delta\NA^2 (  \bar{\boldsymbol \lambda} \wedge  \bar{ \boldsymbol \lambda})
\end{eqnarray}

\section{Inhomogeneous magnetic fields}
\label{app:inhomogeneous}

The microscopic spin operators evolve as
\begin{equation}
{\bf f}^i(t) = R(z_i, t) {\bf f}^i(0) 
\end{equation}
where ${\bf f}^i$ is the spin of the $i$'th atom with position $z_i$ and 
\begin{eqnarray}
R(z, t) &=& \exp[\gamma_F  t |{\bf B}(z)| A_B] 
\end{eqnarray}
where
\begin{eqnarray}
A_B &\equiv&  \left(\begin{array}{ccc} 
0 & -\Bunit_z &\Bunit_y \\ 
 \Bunit_z & 0 & -\Bunit_x \\ 
-\Bunit_y & \Bunit_x & 0  \end{array} \right),
\end{eqnarray} 
is the generator of rotations about ${\bf B}$ and $\hat{\bf B} \equiv {\bf B}/|{\bf B}|$. Expanding the field as 
${\bf B}(z) \approx {\bf B}_0 + (\bBpar + \bBperp )  z + O(z^2)$, where $\bBpar$ is parallel to ${\bf B}_0$ and $\bBperp$ is perpendicular. We note that a change in the magnitude of ${\bf B}$ has an accumulating effect on the spin precession, i.e., the change in ${\bf f}$ grows with $t$.  In contrast, a change in the  {direction} of ${\bf B}$ has a fixed effect:  From the perspective of the measurement, a rotation of ${\bf B}$ is equivalent to a rotation of both the initial state and the measured component $F_z$.  For small gradients $\partial_z {\bf B} \ll {\bf B}/l_{\rm atoms}$, where $l_{\rm atoms}$ is the length of the cloud, we can ignore $\bBperp$.  This approximation, along with the fact that $A_B^{n+2} = - A_B^n$, allows us to write 
\begin{eqnarray}
R(z, t) 
 & \approx & \mathbb{I} + A_{B_0} \sin \omega(z) t + A_{B_0}^2 [1-\cos \omega(z) t ]
\end{eqnarray}
where $\omega(z) = \gamma_F |{\bf B}_0 + z \bBpar|$.

In our trap, we observe an atomic density $\rho(z)$ well approximated by a Lorenzian $\rho(z) =  \width / \pi(z^2 + \width^2)$ where $\width$ is the full-width half-maximum (FWHM) extent of the ensemble. The collective spin ${\bf F} \equiv \sum_i f^i $ then evolves as
\begin{eqnarray}
{\bf F}(t) &=& \int dz \, \rho(z) R(z,t) {\bf f}(0) \\
& = & [\mathbb{I}\, + A_{B_0}^2 ] {\bf F}(0) \nonumber \\
& & + e^{-\width \gamma_F |\bBpar| t} (A_{B_0}\sin  \omega_0 t - A_{B_0}^2\cos \omega_0 t  ) {\bf F}(0) 
\end{eqnarray}
In the first term $\mathbb{I}\, + A_{B_0}^2$ describes a projector onto the direction of ${\bf B}_0$.  This is the steady-state polarization.  The second line describes a decaying oscillation about ${\bf B}_0$ of the perpendicular components with a coherence time $T=1/(\width \gamma_F |\bBpar|)$. 

\end{document}